\documentclass[11pt,preprint,authoryear]{elsarticle}

\makeatletter
\def\ps@pprintTitle{%
 \let\@oddhead\@empty
 \let\@evenhead\@empty
 \def\@oddfoot{\centerline{\thepage}}%
 \let\@evenfoot\@oddfoot}
\makeatother

\usepackage{a4wide}
\usepackage{amsmath}
\usepackage{amsfonts}
\usepackage{amssymb}
\usepackage{graphicx}
\usepackage{hyperref}
\usepackage[toc,page]{appendix} 

\newtheorem{thm}{Theorem}
\newproof{pf}{Proof}
\newtheorem{Rem}{\textit{Remark}}

\newcommand{\argmin}[2]{%
\smash{\mathop{{\rm argmin}}\limits_{#1}}\,#2} 
\newcommand{\argmax}[2]{%
\smash{\mathop{{\rm argmax}}\limits_{#1}}\,#2} 
\def\ee{{\mathbb E}}

\def\rr{{\mathbb R}}

\newcommand{\mb}[1]{\mathbf{#1}}

\makeatletter
\providecommand{\doi}[1]{%
  \begingroup
    \let\bibinfo\@secondoftwo
    \urlstyle{rm}%
    \href{http://dx.doi.org/#1}{%
      doi:\discretionary{}{}{}%
      \nolinkurl{#1}%
    }%
  \endgroup
}
\makeatother

\begin{document}

\begin{frontmatter}

\title{Bayesian inference and non-linear extensions of the CIRCE method for quantifying the uncertainty of closure relationships integrated into thermal-hydraulic system codes}

\author[1]{Guillaume Damblin\corref{cor1}}
\ead{guillaume.damblin@cea.fr}
\cortext[cor1]{Corresponding author}

\author[]{Pierre Gaillard\fnref{fn2}}
\ead{pierre.gaillard@framatome.com}
\fntext[fn2]{The main content of the paper was done when Pierre Gaillard worked in CEA. He is currently working in Framatome. Present address: Framatome - 1, place de la Coupole - Jean-Millier - 92400 Courbevoie, France}

\address[1]{CEA Saclay - DEN/DANS/DM2S/STMF - F-91191 Gif-sur-Yvette Cedex, France}

\begin{abstract}
Uncertainty Quantification of closure relationships integrated into thermal-hydraulic system codes is a critical prerequisite in applying the Best-Estimate Plus Uncertainty (BEPU) methodology for nuclear safety and licensing processes. This issue has been subject to several international initiatives, such as BEMUSE and PREMIUM projects, as well as some statistical developments of which the ``CIRCE" method. This method has been designed at the end of the twentieth century, then extensively used for quantifying the uncertainty of closure relationships integrated into the CATHARE thermal-hydraulic system code. 

\medskip
The purpose of the CIRCE method is to estimate the (log)-Gaussian probability distribution of a multiplicative factor applied to a reference closure relationship in order to assess its uncertainty. Even though this method has been implemented with success in numerous physical scenarios, it can still suffer from substantial limitations such as the linearity assumption and the difficulty of properly taking into account the inherent statistical uncertainty. In the paper, we will extend the CIRCE method in two aspects. On the one hand, we adopt the Bayesian setting putting prior probability distributions on the parameters of the (log)-Gaussian distribution. The posterior distribution of the parameters is then computed with respect to an experimental database by means of Markov Chain Monte Carlo (MCMC) algorithms. 
On the other hand, we tackle the more general setting where the simulations do not move linearly against the multiplicative factor(s). MCMC algorithms then become time-prohibitive when the thermal-hydraulic simulations exceed a few minutes. This handicap is overcome by using Gaussian process (GP) emulators which can yield both reliable and fast predictions of the simulations. 

\medskip
The GP-based MCMC algorithms will be applied to quantify the uncertainty of two condensation closure relationships at a safety injection with respect to a database of experimental tests. The thermal-hydraulic simulations will be run with the CATHARE 2 computer code.
\end{abstract}

\end{frontmatter}

\section{Introduction}

The use of Best-Estimate system codes for nuclear applications in line with the so-called BEPU (Best-Estimate Plus Uncertainty) methodology has been encouraged from several years for reactor transient simulations and safety analysis. This includes code development as well as V\&V (Verification and Validation) and UQ (Uncertainty Quantification) efforts. The paper written by \citet{Dauria12} gives an overview of the best practices for safety demonstration in the nuclear community and how the BEPU approach could meet them. As its name implies, the second stage of a BEPU methodology aims to quantify uncertainties related to the numerical simulation of a thermal-hydraulic transient under study. Such uncertainties can have multiple origins but it is commonly admitted that we can split them into two categories: those of numerical nature (out of the scope of the paper), and those related to the modeling of physical phenomena on which the code relies. The Verification step aims to assess the former, essentially by fixing code bugs and computing discretization errors induced by mesh size. Then, the Validation step aims to quantify the remaining gap between the simulations and the real system by means of some well chosen field experiments used as references \citep{Bayarri2007}. In the last two decades, the research dedicated to V\&V and UQ activities has led to the development of exhaustive guidelines in the field of computational fluid dynamics \citep{AIAA1998,ASME} which have been adapted for reactor safety licensing under the name of improved BEPU \citep{BEPU}. The improved BEPU framework provides general requirements which help improving the seminal methods implemented by thermal-hydraulic experts such as CSAU \citep{CSAU} and GRS approach \citep{Glaeser08}. Both of these methods, developed respectively by the US Nuclear Regulatory Commission (NRC) and the German nuclear safety institution, propagate through the code the input uncertainties to the outputs of interest, such as the Peak Cladding Temperature (PCT). Their implementation for accidental transients at the full reactor scale has been the subject of the BEMUSE\footnote{Best Estimate Methods – Uncertainty and Sensitivity Evaluation} program \citep{Bem08} where LB-LOCA\footnote{Large Break - Loss Of Coolant Accident} analyses were performed to check whether both scaling and uncertainty propagation could be properly implemented. The probability distributions of uncertain input parameters were assumed to be known in BEMUSE because their determination was not the point of the project. This issue has actually been studied in two further international projects named PREMIUM\footnote{Post-BEMUSE REflood Models Input Uncertainty
Methods} \citep{Skorek2019}, following by the recent SAPIUM project\footnote{Systematic Approach for Input Uncertainty Quantification} \citep{Baccou19}. The former aimed to compare in a reflood scenario the performance of several methodologies for quantifying the uncertainty of closure relationships, including, in particular, that of CEA called CIRCE\footnote{In French the acronym ``CIRCE" means \textit{Calcul des Incertitudes Relatives aux Corr\'elations \'El\'ementaires}} and the Fast Fourier Transform-based method \citep{Freixa16}. Unfortunately, the lack of consensus as well as the various practices between the participants raised the need of establishing guidelines and providing general recommendations on how this issue should be addressed. This was the goal of the SAPIUM project.

One of the main source of uncertainty is the one tainting the closure relationships integrated into the balance equations of thermal-hydraulic system codes. Such codes rely on a two phase flow six equations model, consisting of three balance equations for the liquid phase and three for the steam phase (mass, momentum and energy for each phase). The solution of such equations depends on closure relationships constructed from appropriate physical experiments called SETs (Separate Effect Tests). The SETs, often performed at a smaller scale than that of actual nuclear reactors, are instrumented in order to focus on some experimental Quantities of Interests (QoIs) as independently as possible of potential interactions with other phenomena. Then, based on a careful analysis of such tests, the closure relationships under study are established by thermal-hydraulic experts. At the end of this process, though, a model-form uncertainty can remain due to the lack of perfect knowledge of their mathematical expressions. This type of uncertainty can be quantified by the CIRCE method which applies a random multiplicative factor to each uncertain closure relationship \citep{DeCrecy01}. Such factors are in fact modeled either by a Gaussian or log-Gaussian probability distribution whose parameters are estimated from the differences between the experimental QoIs and the corresponding simulations. This estimation stage requires a linear connection between the code outputs and the involved multiplicative factor(s).

In the CIRCE method, the parameters of the (log)-Gaussian distribution are estimated as the Maximum Likelihood Estimator
(MLE), which is the configuration of the parameters at which the probability of observing the data is the highest. 
On the other side, the Bayesian paradigm is to represent uncertain parameters by random variables. First, a prior probability distribution is put on the uncertain parameters, then this distribution is converted into a posterior distribution through the Bayes formula (see for example \citet{Gosh07} for an introduction to Bayesian analysis). Various point estimates, such as the maximum a posteriori and the posterior mean can be computed as well as credible intervals. 
Besides, Bayesian inference is particularly suited for industrial contexts where the prior distribution can benefit from expert judgment, such as accounting for constraints between parameters, upper and lower bounds of variation or even guesses on the best fitting values. In the paper, we will present the Bayesian approach to estimate the parameters of the (log)-Gaussian distributions of the multiplicative factors. Then, the linear assumption of the CIRCE method will be dropped to tackle the case where the simulations move non linearly with respect to the multiplicative factors. However, although thermal-hydraulic simulations are moderately time consuming, intensive sampling required by Bayesian computation makes necessary the use of emulators. We have chosen Gaussian process (GP) emulators which can yield both reliable and fast predictions of the simulations.

\medskip
Section $2$ recalls the foundations of the CIRCE method, and the statistical equation linked to it. Section $3$ deals with Bayesian estimation in the linear setting via Markov Chain Monte Carlo (MCMC) algorithms. Section $4$ goes ahead with the non linear setting and how it modifies MCMC sampling, now based on GP emulators. In Section $5$, the MCMC algorithms are applied to the COSI (Condensation at the Safety Injection) test facilities to quantify the uncertainty of two condensation models at a safety injection. Section $6$ gives conclusions and provides some ideas for future developments.

\section{The CIRCE method}

\label{secCIRCE}

\subsection{Motivation}

\medskip
The CIRCE method is devoted to Uncertainty Quantification (UQ) of physical models integrated into thermal-hydraulic system codes. Such models play the role of closure relationships of the balance equations on which thermal-hydraulic system codes are based. In a BEPU context, the implementation of the CIRCE method is motivated by the requirement of assessing the uncertainty of a few targeted quantities for safety, such as the Peak Cladding Temperature (PCT). This can be done by propagating through the code the uncertainty of the physical models to the PCT (if possible, both scaling and systematic biases should be taken into account as well).
Prior to such a forward UQ stage, physical models are established by experts by means of one or several experimental database(s). The reference formulation of a physical model is written in the paper as
\begin{equation}
M_{ref}(\mb{x})
\end{equation}
with $\mb{x}$ being the vector of the involved thermal-hydraulic conditions such as pressure, temperature, geometrical properties, dimensionless numbers, void fraction, and so on. As such reference models are tainted by some degree of empiricism in their mathematical expressions, the UQ stage is compulsory. To this end, the CIRCE method applies a multiplicative factor $\lambda$ to each reference model of the thermal-hydraulic code. This allows to measure a relative model uncertainty, which is suited to deal with experimental databases having different orders of magnitude. A thermal-hydraulic simulation can then be run with respect to a perturbed model $M_{\lambda}(\mb{x})$ such that
\begin{equation}
M_{\lambda}(\mb{x})=\lambda \times M_{ref}(\mb{x}).
\end{equation}
The more $\lambda$ moves away from $\lambda=1$, the more the engineer should question the accuracy of the reference model. Plausible bounds can sometimes be specified around $\lambda=1$ using expert knowledge. However, the way of representing the uncertainty between the lower and the upper bound often remains unclear. To do so, the CIRCE method adopts a probabilistic framework where the multiplicative factor is modeled by a (log)-Gaussian probability distribution $\Lambda$. When realizations $\lambda\thicksim\Lambda$ are not experimentally observed, the parameters of $\Lambda$ cannot be readily estimated by empirical mean and variance estimators. However, if the impact of $\Lambda$ on some thermal-hydraulic QoIs predicted by the code is strong enough, a statistical modeling of the mismatch between the experimental QoIs and the corresponding simulations at the reference model can be set to get back to the parameters of $\Lambda$. Let us consider, for example, an heat transfer model impacting a wall temperature in a pipe. The differences between the wall temperature measurements and the corresponding simulations could be used to estimate the (log)-Gaussian factor applied to the heat transfer model.

\subsection{The statistical modeling}

\smallskip
As explained previously, a database of experimental QoIs is needed to implement the CIRCE method. Throughout the paper, the superscript $f$ (for \textit{field}) will refer to any experimental quantity. The set-ups of the database are gathered in the matrix $\mb{X}^{f}:$
\begin{equation}
\mb{X}^{f}=[\mb{x}_1^{f},\cdots,\mb{x}_n^{f}]^{T}\in \mathcal{M}_{n,q}(\rr)
\end{equation}
where each site $\mb{x}_i^{f}\in\rr^{q}$ is made up with the physical conditions of the $i$-th experiment ($1\leq i\leq n$). The corresponding QoI at $\mb{x}_i^f$ is denoted by $z_i^{f}\in\rr$. The vector gathering all these QoIs is denoted by
\begin{equation}
\mb{z}^{f}=(z_1^{f},\cdots,z_n^{f})^{T} \in \rr^{n}.
\end{equation}

\smallskip
Let $Y(.)$ be a thermal-hydraulic system code. The CIRCE method relies on a statistical equation which relates $\mb{z}^{f}$ to the corresponding simulations run at $\mb{X}^{f}$. The differences between the two are assumed to be realizations of $n$ independent zero-mean Gaussian random variables. 
Hence, for $1\leq i\leq n$,
\begin{equation}
\label{CIRCEmodel}
z_i^{f}=\, Y(M_{\lambda_i}(\mb{x}_i^{f}))+\epsilon_i  
\end{equation}
with $\lambda_i$ being missing (unobserved) and
\begin{equation}
\label{residual_errors}
\epsilon_i\thicksim\mathcal{N}(0,\sigma_{\epsilon_i}^2).
\end{equation}
The random variable $\epsilon_i$ of Equation (\ref{residual_errors}) is the experimental error (or observation error) whose variance $\sigma_{\epsilon_i}^2$ encompasses the measurement error and a residual variability between the computer code and the physical reality. Other sources of uncertainty could also contribute to the mismatch. In \citet{Koh2001} a function $b(\mb{x}_i^f)$, called "code discrepancy", is explicitly taken into account between the physical system and the computer code, which leads to the equation
\begin{equation}
\label{CIRCEmodel_koh}
z_i^{f}=\, Y(M_{\lambda_i}(\mb{x}_i^{f}))+b(\mb{x}_i^{f})+\epsilon_i  
\end{equation}
Such a modeling has recently be applied to the nuclear field, but considering $\lambda$ as a vector of calibration parameters \citep{Wu18a,Wu18b} (see \ref{Calib_vs_CIRCE} for details on the difference between CIRCE and calibration methods). In the paper, we omit this term because the CIRCE method presumes that the gap between $\mb{z}^{f}$ and the simulations is mostly due to the uncertainty tainting the physical models. Moreover, the inclusion of such a code discrepancy can cause non identifiability problems from a statistical point of view \citep{Bry_OHagan14}.

\medskip
Equation (\ref{CIRCEmodel}) includes the case where the code outputs depend on several uncertain models, $\lambda_i$ being a $p$-dimensional sample of the random vector $\Lambda:=(\Lambda_1,\cdots,\Lambda_p)^{T}$. In such a case, we have
\begin{equation}
\label{Model}
M_{\lambda_i}(\mb{x}_i^{f}):=(M^{1}_{\lambda_{i,1}}(\mb{x}_i^{f}),\cdots,M^{p}_{\lambda_{i,p}}(\mb{x}_i^{f}))
\end{equation} 
where $\lambda_{i,j}$ is the unobserved realization applied to the $j$-th reference model $M^j_{ref}(.)$ ($1\leq j\leq p$). 

\medskip
Equation (\ref{CIRCEmodel}) does not mean a composition of the two functions $Y$ and $M_{\lambda_i}(.)$. In fact, thermal-hydraulic simulations result from an iterative process by which the balance equations are solved in both time and space. The reference model values are modified across iterations whereas the realization $\lambda\thicksim\Lambda$ is constant and should be considered as an extra input value. Hence, Equation (\ref{CIRCEmodel}) can be rewritten under a more readable way:
\begin{equation}
\label{CIRCEmodel_short}
z_i^{f}=\, Y_{\lambda_i}(\mb{x}_i^{f})+\epsilon_i.
\end{equation}

\medskip
The CIRCE method then relies on several important assumptions. First, all the experimental variances $\sigma^2_{\epsilon_i}$ are assumed known ($1\leq i\leq n$). Second, each component $\Lambda_j$ of $\Lambda$ follows either a Gaussian or a log-Gaussian probability distribution$:$
\begin{equation}
\Lambda_j=\mathcal{N}(m_j,\sigma_j^2)\,\,\,\,\,\text{or}\,\,\,\,\,\Lambda_j=\mathcal{LN}(m_j,\sigma_j^2),\,\,\,\,\,\,\,1\leq j\leq p.
\end{equation}
The next assumption is not compulsory to run CIRCE, but highly recommended$:$ the $p$ variables $\Lambda_j$ are independent from one another. For $1\leq j\neq k\leq p$, we set
\begin{equation}
\text{Cov}(\Lambda_j,\Lambda_{k})=0.
\end{equation}
The number of parameters to be estimated is therefore limited to $2p$ instead of $2p+p(p-1)/2$ if cross-covariance terms were estimated as well.
In the sequel, we will use concise notations for the $p$ means and variances of $\Lambda$ to be estimated$:$ 
\begin{equation}
m=(m_1,\cdots,m_p)^{T}\in\rr^{p}
\end{equation}
and 
\begin{equation}
\sigma^2=(\sigma_1^2,\cdots,\sigma_p^2)^{T}\in\rr^{+^p}.
\end{equation}


\subsection{The estimation stage}

\medskip
The estimation of $(m,\sigma^2)$ requires a linear approximation of the code outputs at a nominal value $\lambda^{\star}$ using a first order Taylor approximation.
It is quite frequent that thermal-hydraulic simulations move linearly with respect to the values of the multiplicative factors in a large vicinity of $\lambda^{\star}$. For each input site $\mb{x}_i^f\in\mb{X}^{f}$, Equation (\ref{CIRCEmodel}) is then substituted by 
\begin{equation}
\label{CIRCEmodel_lin_lambda}
z_i^{\prime,f}=\,\, h^{T}(\mb{x}_i^{f})\lambda_i^{\prime}+\epsilon_i
\end{equation}
where 
\begin{itemize}
\item $\lambda_i^{\prime}:=\lambda_i-\lambda^{\star}\thicksim\mathcal{N}(m-\lambda^{\star},\Sigma:=\textrm{diag}(\sigma^2))$ is the missing model sample shifted of $\lambda^{\star}$,
\item $h(\mb{x}_i^{f})\in\rr^{p}$ is the vector of partial derivatives of the code at $\mb{x}_i^{f}$,
\item $z_i^{\prime,f}$ is the difference between the experimental QoI at $\mb{x}_i^f$ and the corresponding simulation at $\lambda^{\star}$.
\end{itemize}
The location $\lambda^{\star}$ is chosen by default equal to $\mb{1}_{p}$, meaning that the linear approximation is constructed at the reference model.

\medskip
In the thermal-hydraulic field, such linear approximations are often more accurate with respect to $\alpha=\log{\lambda}$. Equation (\ref{CIRCEmodel}) is then replaced around $\alpha^{\star}=\log{(\lambda^{\star})}$ by 
\begin{equation}
\label{CIRCEmodel_lin_alpha}
z_i^{\prime,f}=\,\, h^T(\mb{x}_i^{f})\alpha_i^{\prime}+\epsilon_i
\end{equation}
where  
\begin{itemize}
\item $\alpha_i^{\prime}=\alpha_i-\alpha^{\star}\thicksim\mathcal{N}(m-\alpha^{\star},\Sigma:=\textrm{diag}(\sigma^2))$ is the missing model log-sample shifted of $\alpha^{\star}$,
\item $h(\mb{x}_i^{f})\in\rr^{p}$ is the vector of partial derivatives of the code at $\mb{x}_i^{f}$ (with respect to $\alpha=\alpha_{\star}$).
\end{itemize}
The location $\alpha^{\star}$ is chosen by default equal to $\mb{0}_{p}=\log{\mb{1}_{p}}$ meaning that the log-linear approximations are still constructed at the reference model. 

\medskip
Both Models (\ref{CIRCEmodel_lin_lambda}) and (\ref{CIRCEmodel_lin_alpha}) are statistically identifiable if and only if the rank of the corresponding matrix of the partial derivatives matrix is equal to $p$. This matrix is denoted by
\begin{equation}
\label{H}
H=[h(\mb{x}_1^{f}),\cdots,h(\mb{x}_n^{f})]^{T}\in \mathcal{M}_{n,p}(\rr).
\end{equation}
A small condition number is also expected to ensure reliability of estimates \citep{Belsley80}. The Maximum Likelihood Estimate (MLE) of $(m,\sigma^2)$, denoted by $(\hat{m},\hat{\sigma}^2)$, is then computed by means of a variant of the Expectation-Maximization (EM) algorithm called ECME \citep{EM77,Celeux10}. 

\medskip
The last stage of CIRCE, but not the least, is to choose between the Gaussian distribution $\mathcal{N}(\hat{m},\hat{\sigma}^2)$ and the log-Gaussian distribution $\mathcal{LN}(\hat{m},\hat{\sigma}^2)$ where $(\hat{m},\hat{\sigma}^2)$ are the estimates of Models (\ref{CIRCEmodel_lin_lambda}) and (\ref{CIRCEmodel_lin_alpha}) respectively. The choice goes in favor of the distribution making the best linear approximation of the code outputs in the vicinity of $\lambda^{\star}$ or $\alpha^{\star}$ respectively. If the Gaussian distribution is chosen for $\Lambda_j$, then the $95\%$-fluctuation interval of the multiplicative factor is 
\begin{equation}
\label{IFgauss}
IF_{95\%}(\Lambda_j)=[\hat{m}_j-1.96\hat{\sigma}_j,\hat{m}_j+1.96\hat{\sigma}_j]
\end{equation}
Otherwise, if the log-Gaussian distribution is chosen instead, then the $95\%$-fluctuation interval of the multiplicative factor is 
\begin{equation}
\label{IF_log_gauss}
IF_{95\%}(\Lambda_j)=[\exp{(\hat{m}_j-1.96\hat{\sigma}_j)},\exp{(\hat{m}_j+1.96\hat{\sigma}_j)}].
\end{equation}
Either of these intervals is valid if the linearity assumption is well verified across it. Eventually, the probability distribution $\Lambda$ should be propagated through the code to check whether the resulting simulations envelop the physical experiments $\mb{z}^{f}$. This is called \textit{Envelop calculations} in the CIRCE guidelines. Also important is to compare the predicted residuals with the Gaussian density to ensure they broadly match each other (such validation steps of the CIRCE method are out of the scope of the paper).

\begin{Rem}
The statistical uncertainty is assessed with confidence intervals centered on $(\hat{m},\hat{\sigma}^2)$ by using either bootstrapping, or the Fisher information matrix. However, this uncertainty is not taken into account in the computation of the fluctuation intervals.
\end{Rem}

In the next section, we lay out the Bayesian version of the CIRCE method where the uncertain parameters $m$ and $\sigma^2$ are now modeled by random variables. A prior distribution is first put on $(m,\sigma^2)$, then the posterior distribution of $(m,\sigma^2)$ is computed using the Bayes formula. 



\section{The Bayesian inference of the CIRCE method}

\label{Bayes_lin_sec}

\subsection{General principle}

\medskip
Without loss of generality, we present the Bayesian equations related to Model (\ref{CIRCEmodel_lin_alpha}) where $\Lambda$ is specified as a log-Gaussian probability distribution. As the code is linearized at $\alpha^{\star}=\log{\lambda^{\star}}$, we deal with Bayesian estimation of $b:=m-\alpha^{\star}$ and $\sigma^2$. 

\label{gibbs_principle}

\medskip
Let $\mb{A}^{\prime}=[\alpha_1^{\prime},\cdots,\alpha_n^{\prime}]^{T}\in \mathcal{M}_{n,p}(\rr)$ and $\mb{z}^{\prime,f}=(z_1^{\prime,f},\cdots,z_n^{\prime,f})^{T}$ be respectively the matrix of the $\alpha^{\star}$-shifted missing values of the multiplicative factors and the vector of the $\alpha^{\star}$-shifted experimental QoIs. Bayesian estimation of $b$ and $\sigma^2$ begins by specifying a joint prior probability density $\pi(b,\sigma^2)$. Then, applying the Bayes formula gives the posterior density 
\begin{equation}
\label{BFapplication}
\pi(b,\sigma^2|\mb{X}^{f},\mb{z}^{\prime,f})=\frac{\mathcal{L}(\mb{z}^{\prime,f}|\mb{X}^{f},b,\sigma^2)\pi(b,\sigma^2)}{\mathcal{L}(\mb{z}^{\prime,f})}
\end{equation}
where the numerator is the product between the likelihood related to Equation (\ref{CIRCEmodel_lin_alpha}) and the prior distribution $\pi(b,\sigma^2)$, while the denominator is the marginal likelihood of $\mb{z}^{\prime,f}$ playing the role of normalization constant. The likelihood in Equation (\ref{BFapplication}) can be written as
\begin{equation}
\label{margLikeli}
\mathcal{L}(\mb{z}^{\prime,f}|\mb{X}^{f},b,\sigma^2)=\int_{\rr^{n\times p}}\mathcal{L}^{C}(\mb{z}^{\prime,f},\mb{A}^{\prime}|\mb{X}^{f},b,\sigma^2)d \mb{A}^{\prime}
\end{equation}
where $\mathcal{L}^{C}$ is the complete likelihood including the missing data $\mb{A}^{\prime}$. By taking into account these data as unknown quantities as well, the Bayes formula gives
\begin{equation}
\label{BFapplication_v2}
\pi(b,\sigma^2,\mb{A}^{\prime}|\mb{X}^{f},\mb{z}^{\prime,f})\,\propto\, \mathcal{L}(\mb{z}^{\prime,f}|\mb{A}^{\prime},\mb{X}^{f},b,\sigma^2)\pi(\mb{A}^{\prime},b,\sigma^2).
\end{equation}
The prior distribution in Equation (\ref{BFapplication_v2}) can be expanded on
\begin{equation}
\label{prior_hierarchical}
\pi(\mb{A}^{\prime},b,\sigma^2)=\pi(\mb{A}^{\prime}|b,\sigma^2)\pi(b,\sigma^2)
\end{equation}
where $\pi(\mb{A}^{\prime}|b,\sigma^2)$ is the density of the product between $n$ Gaussian distributions having same mean $b$ and same variance $\sigma^2$. Equation (\ref{BFapplication_v2}) shows a hierarchical Bayesian modeling where no distinction is made between the missing data $\mb{A}^{\prime}$ and the parameters $(b,\sigma^2)$, both being unknown and modeled by random variables.
As the computation of the denominator in Equation (\ref{BFapplication}) is challenging, the posterior distribution (\ref{BFapplication_v2}) cannot be readily computed. Instead, an efficient way to generate samples from this distribution is to run Markov Chain Monte Carlo (MCMC) algorithms relying on the numerator only.

\subsection{Computation of the full conditional posterior distributions}

In such a linear setting, MCMC methods can be used to generate samples from the posterior distribution (\ref{BFapplication_v2}). MCMC methods are able to generate correlated samples of intractable probability distributions. The algorithm we present below is referred to as the blocked Gibbs sampler, which was originally called substitution sampling \citep{Gelfand90}. Its sampling scheme follows the prior structure in Equation (\ref{prior_hierarchical}), but now conditional on experimental data $(\mb{z}^{\prime,f}, \mb{X}^f)$.

\bigskip
\vspace{3pt}\hrule\vspace{6pt}
\noindent \textbf{Blocked Gibbs sampler} 
\vspace{3pt}\hrule\vspace{6pt}
\begin{enumerate}[1.]
\item Set starting values: $(b(0),\sigma^2(0))$
\item Choose a number $N_{mcmc}$ of iterations, then for $1\leq k\leq N_{mcmc}$, generate in a loop:
\begin{equation}
\label{fcd1_blocked_main}
\mb{A}^{\prime}(k+1) \thicksim \mb{A}^{\prime}\,|\, b(k),\sigma^2(k),\mb{z}^{\prime,f}, \mb{X}^f
\end{equation}
\begin{equation}
\label{fcd2_blocked_main}
b(k+1),\sigma^2(k+1)\thicksim b,\sigma^2 \,|\, \mb{A}^{\prime}(k+1),\mb{z}^{\prime,f}, \mb{X}^f
\end{equation}
\end{enumerate}
\vspace{3pt}\hrule\vspace{6pt}

\bigskip
If the conjugate Gaussian-inverse gamma prior distribution is put on $(b,\sigma^2)$, then the distributions (\ref{fcd1_blocked_main}) and (\ref{fcd2_blocked_main}) are analytically tractable (see \ref{MCMC_several_lin} for the proof). Such a prior is defined by
\begin{equation}
\label{inv_gamm_prior}
\pi(b,\sigma^2)=\prod_{j=1}^{p}\pi(b_j,\sigma^2_j)=\prod_{j=1}^{p}\pi(b_j|\sigma_j^2)\pi(\sigma_j^2)
\end{equation}
where for $1\leq j\leq p$
\begin{equation}
\pi(b_j|\sigma_j^2)=\frac{\sqrt{a_j}}{\sqrt{2\pi}\sigma_j}\exp{\Big\{-\frac{a_j(b_j-\mu_j)^2}{2\sigma_j^2}\Big\}}
\end{equation}
and
\begin{equation}
\pi(\sigma_j^2)=\frac{\gamma_j^{\psi_j}}{\Gamma(\psi_j)}\sigma_j^{2(-\psi_j-1)}\exp{\Big\{-\frac{\gamma_j}{\sigma_j^2}\Big\}}.
\end{equation}
The Greek letter $\Gamma(.)$ refers to the Euler gamma function. The hyperparameters $\psi_j,\gamma_j,\mu_j,a_j$ are set depending on the amount of prior information on $(b,\sigma^2)$. If a data-dominated posterior distribution is expected, typically in absence of prior information on $(b,\sigma^2)$, the values $\mu_j=0$, $a_j=\epsilon$, $\psi_j=\epsilon$ and $\gamma_j=\epsilon$ with $\epsilon$ being close to $0$ should be used \citep{Spig04}. However, a non negligible impact of $\epsilon$ on the shape of the posterior distribution may remain when $\mb{z}^{\prime,f}$ is of moderate size \citep{Gelman06}. 

\medskip
The blocked Gibbs algorithm generates samples  $\{\mb{A}^{\prime}(k),(b(k),\sigma^{2}(k))\}_k$ by a Markov chain process whose stationary distribution matches the posterior distribution (\ref{BFapplication_v2}). However, the practical convergence of such MCMC algorithms is often not obvious and need be checked (see \ref{MCMCconv} for a look on this issue). Posterior samples $\{(b(k),\sigma^{2}(k))\}_k$ can be used for computing point estimates of the posterior distribution of $(b,\sigma^2)$ as well as posterior credible intervals. The posterior mean is 
\begin{equation}
\label{postmean}
(\hat{b},\hat{\sigma}^2)=\ee[b,\sigma^2|\mb{X}^{f},\mb{z}^{\prime,f}].
\end{equation}
Another much used point estimate is the Maximum A Posteriori (MAP)$:$ 
\begin{equation}
\label{map}
(\hat{b},\hat{\sigma}^2)=\argmax{b,\sigma^2}{\pi(b,\sigma^2|\mb{X}^{f},\mb{z}^{\prime,f})}.
\end{equation} 
Eventually, posterior samples $\{m(k)\}_k$ can be obtained from the posterior samples $\{b(k)\}_k$ by moving them from $\alpha^{\star}$.

\medskip
\begin{Rem}
Other MCMC algorithms than the blocked Gibbs sampler can be implemented to compute the posterior distribution of $(b,\sigma^2)$ (see \ref{MCMC_several_lin} for statistical details). 
\end{Rem}

\subsection{Bayesian fluctuation intervals}

\label{ModelMarg}

\medskip
A nice property of the Bayesian framework is that the statistical uncertainty of $(m,\sigma^2)$ can be taken into account in computing the fluctuation intervals of the multiplicative factors, which makes them more conservative than those computed by the standard version of CIRCE. Such intervals are derived from the posterior predictive distribution of $\Lambda=(\Lambda_1,\cdots,\Lambda_p)$ whose density is computed as
\begin{equation}
\label{Model_marginal}
\pi(\Lambda|\mb{X}^{f},\mb{z}^{\prime,f})=\int \pi(\Lambda|m,\sigma^2)\pi(m,\sigma^2|\mb{X}^{f},\mb{z}^{\prime,f})\text{d} m\, \text{d} \sigma^2.
\end{equation}
where 
\begin{equation}
\pi(\Lambda|m,\sigma^2)=\prod_{j=1}^{p}\pi(\Lambda_j|m_j,\sigma_j^2)
\end{equation}
with $\pi(\Lambda_j|m_j,\sigma_j^2)$ being the log-Gaussian density of $\Lambda_j$ ($1\leq j\leq p$). 

\medskip
The predictive density $\pi(\Lambda|\mb{X}^{f},\mb{z}^{\prime,f})$ has no closed-form but Monte Carlo simulation can be used to generate samples of it. Let $N$ be the number of samples we want to generate. Then, in a loop, for $1\leq k\leq N$ :
\begin{enumerate}
\item pick up a posterior sample $m(k),\sigma^2(k)\thicksim m,\sigma^2|\mb{X}^{f},\mb{z}^{\prime,f}$,
\item generate a realization $\lambda(k)\thicksim \Lambda| m(k),\sigma^2(k)$.
\end{enumerate}
Such a simulation scheme yields samples $\lambda_j(1),\lambda_j(2),\cdots,\lambda_j(N)$ to be used for kernel density estimation of (\ref{Model_marginal}) as well as computing empirical quantiles of $\Lambda_j$ ($1\leq j\leq p$). Let $q_{0.025}(\Lambda_j)$ and $q_{0.975}(\Lambda_j)$ be respectively the $2.5\%$ and $97.5\%$ empirical quantiles of $\Lambda_j$. The resulting $95\%$-fluctuation interval of $\Lambda_j$ is equal to
\begin{equation}
IF_{95\%}(\Lambda_j)=[q_{0.025}(\Lambda_j),q_{0.975}(\Lambda_j)].
\end{equation}

\subsection{Main limitation of the linear approach}

\label{limitation_linear}

\medskip
On the one hand, the implementation of CIRCE (and its Bayesian version) can be subject to a strong user effect in constructing the matrix $H$ of the partial derivatives in Equation (\ref{H}). As thermal-hydraulic system codes tend to switch from a flow regime (or thermal regime) to another, the values of the finite differences can be strongly impacted by the selected increment. On the other hand, when the size of $\mb{z}^{f}$ is moderate ($30\leq n\leq 100$), the resulting fluctuation intervals are likely to be very spread out. Hence, the linear approximations may fail to match the code outputs for realizations $\alpha$ (resp. $\lambda$) not close enough to $\alpha^{\star}$ (resp. $\lambda^{\star}$). Both of these limitations motivate us to tackle the general setting where the code outputs move non linearly against $\alpha$.

\section{The non linear setting}

\label{nonlin_sec}

\medskip
Let $\mb{A}=[\alpha_1,\cdots,\alpha_n]^{T}\in \mathcal{M}_{n,p}(\rr)$ be the matrix of the missing log-samples of the multiplicative factors. 
As there is no longer linearization of the code outputs, we then revert to the estimation of $m$. The Bayes formula applied to Equation (\ref{CIRCEmodel_short}) with $\lambda=\exp{\alpha}$ gives
\begin{equation}
\label{post_expression_nonlinear}
\pi(m,\sigma^2,\mb{A}|\mb{X}^{f},\mb{z}^{f})\propto \mathcal{L}(\mb{z}^{f}|\mb{A},\mb{X}^{f},m,\sigma^2)\pi(\mb{A}|m,\sigma^2)\pi(m,\sigma^2)
\end{equation}
where $\pi(m,\sigma^2)$ can still be chosen as a Gaussian-inverse gamma prior density. Then, a blocked Gibbs sampler can be run in a similar way to the linear case.

\medskip
\vspace{3pt}\hrule\vspace{6pt}
\noindent \textbf{Blocked Gibbs sampler} (non-linear version)
\vspace{3pt}\hrule\vspace{6pt}
\begin{enumerate}[1.]
\item Set starting values: $(m(0),\sigma^2(0))$
\item Choose a number $N_{mcmc}$ of iterations, then for $1\leq k\leq N_{mcmc}$ generate:
\begin{equation}
\label{fcd1_blocked_nl}
\mb{A}(k+1) \thicksim \mb{A}\,|\, m(k),\sigma^2(k),\mb{X}^f,\mb{z}^{f}
\end{equation}
\begin{equation}
\label{fcd2_blocked_nl}
m(k+1),\sigma^2(k+1)\thicksim m,\sigma^2 \,|\, \mb{A}(k+1),\mb{X}^f,\mb{z}^{f}
\end{equation}
\end{enumerate}
\vspace{3pt}\hrule\vspace{6pt}

\bigskip
The conditional distribution (\ref{fcd2_blocked_nl}) is unchanged compared to the linear case (except we estimate $m$ instead of $b$) whereas the probability density of (\ref{fcd1_blocked_nl}) is now written as
\begin{align}
\nonumber
\pi(\mb{A} | m,\sigma^2,\mb{X}^{f},\mb{z}^{f})=&\,\prod_{i=1}^{n}\pi(\alpha_i | m,\sigma^2,\mb{x}_i^f,z_i^{f}) \\
 \propto & \,\exp{\left\{-\frac{1}{2}\sum_{i=1}^{n}\Big[\frac{\big(z_i^f-Y_{\exp{\alpha_i}}(\mb{x}_i^{f})\big)^2}{\sigma_{\epsilon_i}^{2}}+(\alpha_i-m)^{T}\Sigma^{-1}(\alpha_i-m)\Big]\right\}}
 \label{cond_latent}
\end{align}
which depends on the code outputs $Y_{\exp{\alpha_i}}(\mb{x}_i^{f})$ for $1\leq i\leq n$. This distribution has no longer closed-form and then MCMC methods such as a Metropolis-Hastings (MH) algorithm can be used for sampling from it \citep{metropolis+53} (see \ref{MH}).

\medskip
The more time consuming the simulations are, the more time consuming the blocked Gibbs sampler will be. Because several ten of thousands of iterations are commonly required for this sampler to converge to the posterior distribution, then running it in a reasonable time, say, up to one day, is no longer feasible when the simulations require a couple of minutes or more. This handicap can be circumvented by replacing the code outputs with emulator predictions in Equation (\ref{cond_latent}). We will use the Gaussian process (GP) emulator which can deliver fast predictions of time demanding simulations as well as uncertainty bounds around them \citep{Sacks89,San03}.

\subsection{The Gaussian Process (GP) emulator}

\label{The_GP}

\medskip
Let the computer code output $Y_{\lambda}(\mb{x}_i^f)$ be a function of $\lambda\in\mathcal{L}$ with $\mathcal{L}\subset\rr^{p}$. When simulations are not instantaneous, the code can be considered as an unknown quantity. Hence, from a Bayesian point of view, we can put assume the functional prior 
\begin{equation}
\label{gp_definition}
Y_{\mb{x}_i^f}(\lambda)\,\,\,\,\,\thicksim\,\,\,\,\, m_{\mb{x}^f_i}(\lambda)+\mathcal{GP}\big(0,K_{\mb{x}_i^f}(\lambda,\lambda^{\prime})\big)
\end{equation}
as the sum of a deterministic and a stochastic part being a second order stationary Gaussian Process (GP).
The positions of $\lambda$ and $\mb{x}_i^{f}$ are reversed in Equation (\ref{gp_definition}) to point out that the GP is a random function of $\lambda\in\mathcal{L}$ at a fixed site $\mb{x}_i^f$. Let $\mb{D}_M$ be a design of $M$ locations of the multiplicative factors$:$
\begin{equation}
\mb{D}_M=[\lambda_1,\cdots,\lambda_M]^{T}\in \mathcal{M}_{M,p}(\rr)
\end{equation}
The corresponding set of simulations run at $\mb{x}_i^f$ is denoted by $Y_{\mb{x}_i^f}(\mb{D}_M)$. The prediction of the code output at a new location $\lambda\in\mathcal{L}$ is given by the random process conditional on $Y_{\mb{x}_i^f}(\mb{D}_M)$, which is still Gaussian$:$
\begin{equation}
\label{gp_pred}
Y^{M}_{\mb{x}_i^{f}}(\lambda):=Y_{\mb{x}_i^{f}}(\lambda)|Y_{\mb{x}_i^f}(\mb{D}_M)\thicksim\mathcal{GP}\Big(\mu_{\mb{x}_i^f}^{M}(\lambda),V_{\mb{x}_i^f}^{M}\big(\lambda,\lambda^{\prime}\big)\Big)
\end{equation}
where the two terms in the brackets are the predictive mean and variance respectively. Their mathematical expressions are given in \ref{GPappendix}. Equation (\ref{gp_pred}) can be referred to as the posterior GP \citep{Currin91}. The mean function in Equation (\ref{gp_definition}) is commonly assumed as either constant 
\begin{equation}
\label{constantmean}
\beta_i:=\beta_{\mb{x}_i^f}\in\rr
\end{equation}
or a linear function as the product between a vector of regression functions and regression parameters $\beta_i:=\beta_{\mb{x}_i^f}\in\rr^{m}$ of regression parameters$:$
\begin{equation}
m_{\mb{x}^f_i,\beta_i}(\lambda)=H^{T}_i(\lambda)\beta_i.
\end{equation}
The covariance function of a second order stationary GP is decomposed into a scale parameter $\sigma_i^2:=\sigma^2(\mb{x}_i^f)\in\rr^{+}$ multiplied by a correlation function indexed by a vector of correlation lengths $l_i:=l(\mb{x}_i^f)\in\rr^{p}$$:$
\begin{equation}
K_{\mb{x}^f_i,\sigma_i^2,l_i}(\lambda,\lambda^{\prime})=\sigma_i^2 C_{\mb{x}^f_i,l_i}(\lambda,\lambda^{\prime}).
\end{equation}
The most commonly used correlation functions are those of Mat\'ern type due to their flexibility. Let us define $\delta \lambda=|\lambda-\lambda^{\prime}|$ with $\lambda, \lambda^{\prime} \in \mathcal{L}\subset \rr$ and 
$\Gamma(.)$ and $K_{\nu}(.)$ being the Gamma and modified Bessel function of the second kind respectively. The one dimensional Mat\'ern function is written as
\begin{equation}
\label{Matern} 
C^{1d,\nu}_{\mb{x}_i^f,l_i}(\delta\lambda)=\frac{2}{\Gamma(\nu)}\Big( \sqrt{\nu}\frac{\delta\lambda}{l_i} \Big)^{\nu} K_{\nu}\big(2\sqrt{\nu}\frac{\delta\lambda}{l_i}\big)
\end{equation}
with $\nu$ being a hyperparameter setting the regularity of the GP's trajectories. This function quantifies to which extend the code outputs $Y_{\mb{x}_i^f}(\lambda)$ and to $Y_{\mb{x}_i^f}(\lambda^{\prime})$ are related to each other according to the deviation $\delta \lambda$. The higher $\nu$, the smoother the GP trajectories. When $p\geq2$, a multidimensional Mat\'ern function can rely on the deviation $\delta\lambda$ below$:$
\begin{equation}
\delta\lambda=\sqrt{\Big[\sum_{j=1}^{p}\big(\frac{\delta\lambda_j}{l_{i,j}}\big)^{2}\Big]}
\label{isoM}
\end{equation}
with $\lambda=(\lambda_1,\cdots,\lambda_p)^{T}$ and $l_i=(l_{i,1},\cdots,l_{i,p})^{T}$ being respectively the vectors of the $p$ deviations in each dimension and $p$ correlation lengths.
Then, an isotropic Mat\'ern function is constructed by replacing $\delta\lambda$ by (\ref{isoM}) into Equation (\ref{Matern}), which gives
\begin{equation}
\label{MaternMult} 
C_{\mb{x}_i^f,l_i}^{\nu}(\delta\lambda)=C_{\mb{x}_i^f,l_i}^{1d,\nu}\Bigg(\sqrt{\Big[\sum_{j=1}^{p}\big(\frac{\delta\lambda_j}{l_{i,j}}\big)^{2}\Big]}\Bigg).
\end{equation}
In real problems, $\nu$ should be properly selected according to the smoothness of the code outputs. The hyperparameters $\beta_i$, $\sigma_i^2$ and $l_i$ will be estimated from $Y_{\mb{x}_i^{f}}(\mb{D}_M)$ as part of the theorem given in the next section.  

\subsection{The GP-based MH within Gibbs sampler}

\label{The_GP-based_MH}

\bigskip
The following theorem rewrites the probability density (\ref{cond_latent}) by using GP emulation.

\begin{thm}
\label{GPmodular_fcd}
Let us assume that one GP emulator is constructed per site $\mb{x}_i^f\in\mb{X}^{f}$ for $1\leq i\leq n$. The whole set of simulations used for fitting all the GP emulators is denoted by
\begin{equation}
Y(\mb{D}_M):=\cup_{i=1}^{n}Y_{\mb{x}_i^f}(\mb{D}_M). 
\end{equation}
Based on every GP at $\mb{x}_i^f\in\mb{X}^{f}$ given by Equation (\ref{gp_pred}), the so-called GP-based approximation of the probability density (\ref{cond_latent}) is written as
\begin{multline}
\label{cond_latent_pg}
\pi(\mb{A}|Y(\mb{D}_M),m,\sigma^2,\mb{z}^{f}) \propto \prod_{i=1}^{n} \frac{1}{\sqrt{\sigma^2_{\epsilon_i}+V^{M}_{\mb{x}_i^f,\hat{\theta}_i}(\exp{(\alpha_i)},\exp{(\alpha_i}))}}\times \\
\exp{\Big\{-\frac{1}{2}\sum_{i=1}^{n}\Big[(z_i^f-\mu^{M}_{\mb{x}_i^f,\hat{\theta}_i}(\exp{(\alpha_i)})^{T}\big(\sigma^2_{\epsilon_i}+V^{M}_{\mb{x}_i^f,\hat{\theta}_i}(\exp{(\alpha_i)},\exp{(\alpha_i}))\big)^{-1}} \\
(z_i^f-\mu^{M}_{\mb{x}_i^f,\hat{\theta}_i}(\exp{(\alpha_i)})+ 
(\alpha_i-m)^{T}\Sigma^{-1}(\alpha_i-m)\Big]\Big\}
\end{multline}
\end{thm}
with $\hat{\theta}_i:=(\hat{\beta}_i,\hat{\sigma}_i^2,\hat{l}_i)$ being a point estimate of the hyperparameters of the predictive mean and variance.
\begin{pf}
A Bayesian modular approach is implemented in the spirit of \citet{Bayarri2007} (see \ref{modularAppendix} for details).
\end{pf}

This theorem provides an approximation of the density (\ref{cond_latent}) in which the thermal-hydraulic simulations are substituted for the predictive means and variances of the posterior GPs.
The interest of using the probability density (\ref{cond_latent_pg}) in comparison to (\ref{cond_latent}) is that the computer code outputs are replaced by fast-to-evaluate Gaussian predictions. The density (\ref{cond_latent_pg}) should be interpreted as the natural GP-based emulation of (\ref{cond_latent}) because, on the one hand, the code outputs are simply replaced by the predictive means of the GPs and, on the other, the predictive variances are added up to the experimental variances so that the potential departure between the predictive means and the actual code outputs is taken into account. The mismatch between $\mb{z}^f$ and the predictive means of the GPs then consists of both the experimental and the emulator uncertainties.

\medskip
The MH step is now much sped up in sampling (\ref{cond_latent_pg}) instead of (\ref{cond_latent}).
The resulting GP-based MH within Gibbs algorithm consists of sampling the two conditional distributions (\ref{cond_latent_pg}) and (\ref{fcd2_blocked_nl}) one after the other. The more the GP emulators can yield accurate and reliable predictions of the thermal-hydraulic simulations, the closer (\ref{cond_latent_pg}) is expected to (\ref{cond_latent}) and, therefore, the closer the GP-based posterior distribution is expected to the exact posterior distribution (\ref{post_expression_nonlinear}).

\begin{Rem}
\label{closenessGP_exact}
If the predictive variances of the GP emulators are larger than or comparable to the experimental variances $\sigma_{\epsilon_i}^{2}$, then the GP-based posterior distribution can hardly match the exact posterior distribution.
\end{Rem}

\medskip
In the next section, we present the COndensation at the Safety Injections (COSI) experimental database from which the uncertainty of two condensation models at a safety injection will be quantified. 

\section{Application to COSI tests}

\label{cositests}

\subsection{Description}

\medskip
The COSI database is established from a SET facility which reproduces the cold leg and the emergency core cooling injection of a nuclear power plant with a power and volumic scale of 1/100. This facility allows to characterize two heat transfer phenomena to which the two multiplicative factors $\Lambda_1$ and $\Lambda_2$ are applied$:$ 
\begin{itemize}
\item $\Lambda_1$ is applied to the \underline{condensation created on the free surface} far from the injection \citep{Bestion93},
\item $\Lambda_2$ is applied to the \underline{condensation due to the turbulent mixing} in the vicinity of the injection jet \citep{Gaillard16}.
\end{itemize}
The COSI facility is composed of$:$
\begin{itemize}
\item[$\bullet$] a cold leg,
\item[$\bullet$] an emergency core cooling injection,
\item[$\bullet$] a downcomer,
\item[$\bullet$] a boiler to regulate the pressure in the test facility.
\end{itemize}
We have considered steady-state tests. Various thermal-hydraulic input conditions $\{\mb{x}^f_i\in\rr^{4}\}_i$ have been tested, such that$:$
\begin{itemize}
\item[$\bullet$] water height in the cold leg is between 0 and 60\%,
\item[$\bullet$] injection temperature is between 20 and 80 degrees Celsius,
\item[$\bullet$] pressure in the test facility is between 20 and 70 bars,
\item[$\bullet$] injection flow rate is between 0.1 and 0.6 kg/s.
\end{itemize}
The instrumentation has been composed by thermocouples in the cold leg as well as a boiler flow rate measuring device to follow the temperature evolution in the liquid layer and the condensation flow rate in the facility respectively. 

\medskip
The vector $\mb{z}^f$ is composed by $n=50$ condensation flow rate values with respect to various conditions $\mb{x}_i^f$ ($1\leq i\leq n$). The standard deviation of $z_i^f$ is equal to 
$\sigma_{\epsilon_i}=10^{-4}$ for all experiments. The corresponding simulations run with the CATHARE 2 code need a couple of minutes.

\subsection{Preliminary estimations using the standard CIRCE method}

\medskip
Both Gaussian and log-Gaussian parametrizations were compared to each other (see Equation \ref{CIRCEmodel_lin_lambda} and \ref{CIRCEmodel_lin_alpha} respectively). The latter gave the best approximation of the code outputs and then $\Lambda_1$ and  $\Lambda_2$ are log-Gaussian distributions. Applying the CIRCE method yielded the following biases$:$
\begin{equation}
(\hat{b}_1,\hat{b}_2)=(0.85,-0.78).
\end{equation} 
As these values are far away from $0$, the linear approximations made at $\alpha^{\star}=(0,0)$ were poor in the vicinity of $(\hat{b}_1,\hat{b}_2)$. A way of improving the robustness of the CIRCE method is to make successive linearizations until small biases are obtained. This is referred to as the iterative CIRCE method \citep{Nouy17}. For the COSI tests, we carried out four iterations, the last being at $\alpha^{\star}=(\log{1.73},\log{0.48})=(0.55,-0.73)$. The biases computed by CIRCE at this $\alpha^{\star}$ were equal to
\begin{equation}
(\hat{b}_1,\hat{b}_2)=(0.008,-0.019).
\end{equation} 
Equation (\ref{IF_log_gauss}) then gave the resulting $95\%$-fluctuation intervals of $\Lambda_1$ and $\Lambda_2$$:$
\begin{equation}
\label{CirceIF}
IF_{95\%}(\Lambda_1)=[0.84,3.61]\,\,\,\,\,\text{;}\,\,\,\,\,IF_{95\%}(\Lambda_2)=[0.11,1.98]
\end{equation}

\subsection{Bayesian estimation using the linear approximations}

\label{linear_cosi}

\bigskip
By using a diffuse Gaussian-inverse gamma prior distribution with $\epsilon=10^{-2}$, the blocked Gibbs sampler presented in Section \ref{Bayes_lin_sec} was run through an executable C++ macro. We generated $3\times 10^6$ MCMC samples $\{(b(k),\sigma^2(k))\}$ in some minutes, then stored into a \textit{TTree} object of the TDataServer class of the Uranie platform \citep{Blanchard19}. 

\begin{figure}[ht!]
\caption{\textit{Posterior distribution of $(\exp{(m_1)},\exp{(m_2)})$. This figure displays $2.7\times 10^{4}$ quasi-independent samples (obtained after thinning of the full chain to reduce the autocorrelation).}}
\centering
\vspace{0.5cm}
\includegraphics[scale=0.50]{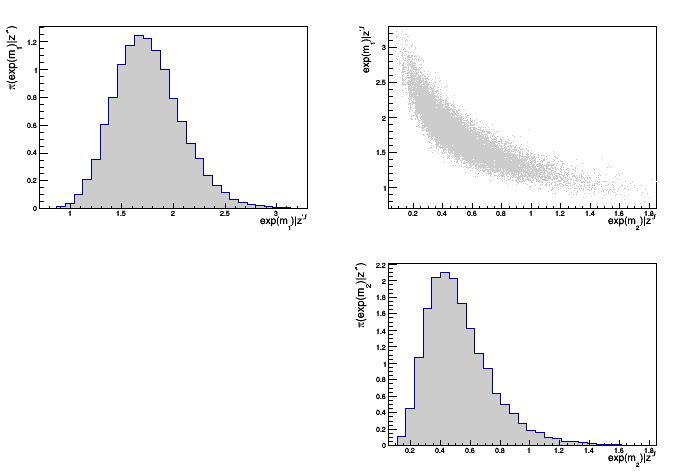}
\label{posterior_distributions_linear_m}
\end{figure}

\begin{figure}[ht!]
\caption{\textit{Posterior distribution of $(\sigma^2_1,\sigma^2_2)$. This figure displays $2.7\times 10^{4}$ quasi-independent samples (obtained after thinning of the full chain to reduce the autocorrelation).}}
\centering
\vspace{0.5cm}
\includegraphics[scale=0.50]{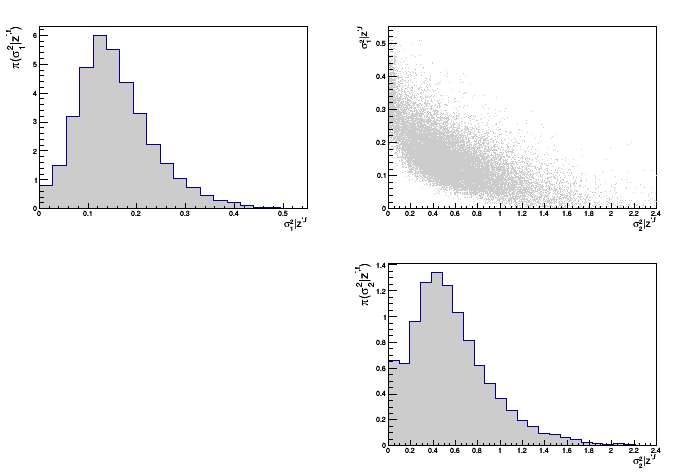}
\label{posterior_distributions_linear_v}
\end{figure}

The Geweke testing was applied to determine the number of iterations, called burn-in period before the chain really converges to the posterior distribution (see \ref{MCMCconv}). As the four $p$ values for $b_1$, $b_2$, $\sigma^{2}_1$ and $\sigma^{2}_2$ were all above $0.05$, no evidence of non convergence was detected in the first part at this level of significance. As a result, we removed the first $10\%$ samples and kept the remaining $90\%$, i.e. $2.7\times 10^6$ samples. Then, we got back to the posterior distribution of $(m_1,m_2)$ by moving of $\alpha^{\star}=(0.55,-0.73)$ the posterior samples of $(b_1,b_2)$. The corresponding posterior means of $m_1$ and $m_2$ were equal to 
\begin{equation}
\hat{m}_1=0.5447\,\,\,\,\,\,\,\text{and}\,\,\,\,\,\,\,\hat{m}_2=-0.7208. 
\end{equation}
implying
\begin{equation}
\exp{(\hat{m}_1)}=1.7241\,\,\,\,\,\,\,\text{and}\,\,\,\,\,\,\,\exp{(\hat{m}_2)}=0.4864. 
\end{equation}
Neither $\exp{(\hat{m}_1)}$ nor $\exp{(\hat{m}_2)}$ is close to $1$. The log-factor $\log{\Lambda_1}$ is thus positively biased whereas $\log{\Lambda}_2$ is negatively.
The posterior means of $\sigma^2_1$ and $\sigma^2_2$ were equal to
\begin{equation}
\hat{\sigma}^2_1=0.1577\,\,\,\,\,\,\,\text{and}\,\,\,\,\,\,\,\hat{\sigma}^2_2=0.5651. 
\end{equation}
The effective sample size (e.s.s.) values were computed from the posterior samples of $m_1$, $m_2$, $\sigma_1^2$, $\sigma_2^2$ (see \ref{MCMCconv}). The e.s.s. values allowed us to compute the $95\%$ confidence intervals of these parameters$:$
\begin{equation}
IC_{95\%}(\hat{m}_1)=[0.5427,0.5467]\,\,\,\,\,\text{;}\,\,\,\,\,IC_{95\%}(\hat{m}_2)=[-0.7253,-0.7164]
\end{equation}
and
\begin{equation}
IC_{95\%}(\hat{\sigma}_1^2)=[0.1569,0.1584]\,\,\,\,\,\text{;}\,\,\,\,\,IC_{95\%}(\hat{\sigma}^2_2)=[0.5618,0.5684]
\end{equation}
Even though the autocorrelation between consecutive samples was moderate, only every $100$-th sample was picked up to present the histograms of $(\exp{(m_1)},\exp{(m_2)})$ and $(\sigma_1^2,\sigma_2^2)$, on Figures \ref{posterior_distributions_linear_m} and \ref{posterior_distributions_linear_v} respectively. The strong negative dependence between $m_1$ and $m_2$ means that a larger median for $\Lambda_1$ can be counterbalanced by a smaller one for $\Lambda_2$, which is in line with simulation of the COSI tests.

\medskip
To finish, the predictive distribution of $\Lambda=(\Lambda_1,\Lambda_2)$ was sampled by the Monte Carlo algorithm explained in Section \ref{ModelMarg} in order to deduce the $95\%$-fluctuation intervals of $\Lambda_1$ and $\Lambda_2$$:$
\begin{equation}
\label{BayesIF}
IF_{95\%}(\Lambda_1)=[0.69,3.97]\,\,\,\,\,\text{;}\,\,\,\,\,IF_{95\%}(\Lambda_2)=[0.08,2.43]
\end{equation}
These intervals are more spread out than those given by Equation (\ref{CirceIF}). This is because the uncertainty of $(m,\sigma^2)$ is fully taken into account in the Bayesian computation. However, we put  into question their validity because the accordance between the linear approximations and the actual code outputs was quite poor in the tails of the intervals (\ref{BayesIF}) (see Section \ref{limitation_linear}). This motivated to redo such a statistical analysis via the non linear setting to give more credence to the fluctuation intervals. 


\subsection{Bayesian estimation in the non linear setting}  

\label{nonlin_cosi}

\medskip

As seen in Section \ref{nonlin_sec}, the GP-based MH within Gibbs sampler requires fitting one GP emulator per input site $\mb{x}_i^f\in \mb{X}^{f}$ by means of the learning simulations $Y_{\mb{x}_i^f}(\mb{D}_M)$ ($1\leq i\leq 50$). 

\begin{figure}[ht!]
\caption{\textit{Red dots $\mb{D}_{M=500}$ are locations used for learning the GP emulators. Blue stars $\mb{D}^{test}_{M=500}$ are locations used for testing the reliability of predictions over $\mathcal{L}=(0,5)\times (0,5)$.}}
\centering
\vspace{0.5cm}
\includegraphics[scale=0.55]{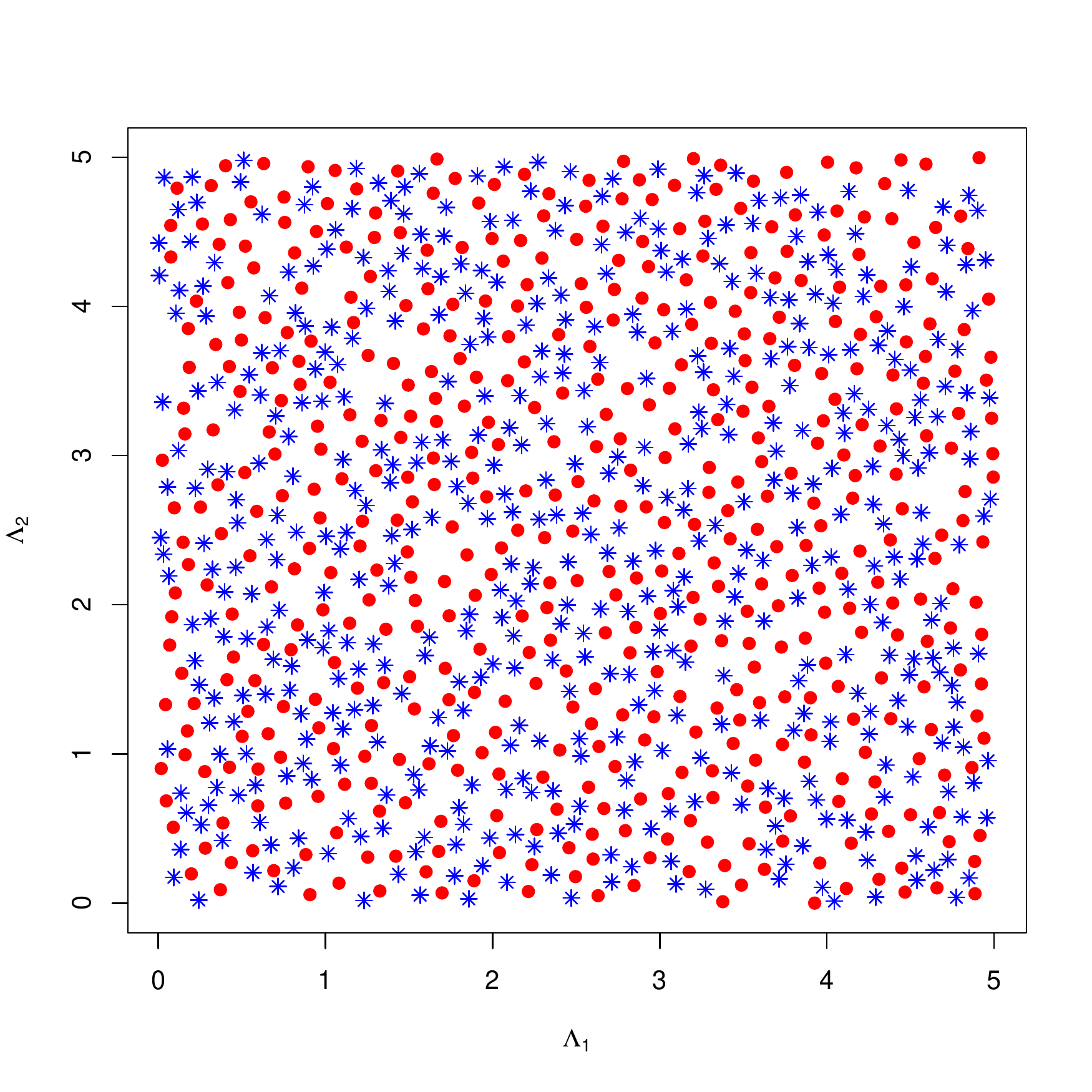}
\label{designs}
\end{figure}

\medskip
The method we applied to construct $\mb{D}_M$ as well as a testing design $\mb{D}^{test}_M$ consists of the three steps below$:$
\begin{enumerate}[1.]
\item construction of a design $\mb{D}_{2M}$ as a Space Filling Design (SFD) with $2M$ locations $\lambda\in\mathcal{L}$. The method \textit{maximinSA\_LHS} from the Dice Design R library was used for doing so \citep{DiceDesign15};

\smallskip
\item construction of $\mb{D}_M$ by using the \textit{wspDesign} method from the same library. This method can take $M$ locations out of $\mb{D}_{2M}$ so that the $M$ remaining ones form a SFD of $\mathcal{L}$; 

\smallskip
\item construction of a testing design $\mb{D}_{M}^{test}=\mb{D}\setminus\mb{D}_M$ to check the reliability of GP predictions.
\begin{equation}
\mb{D}^{test}_M=[\lambda_1^{test},\cdots,\lambda_M^{test}]^{T}\in \mathcal{M}_{M,p}(\rr).
\end{equation}
\end{enumerate}
The choice of $\mathcal{L}$ depends on the values taken by the two multiplicative factors. Based on physical expertise on the two condensation models we chose $\mathcal{L}:=(0,5)\times (0,5)$. Figure \ref{designs} shows both the learning and testing designs with $M=500$ locations for each.

\begin{figure}[ht!]
\caption{\textit{Distribution over $\mb{X}^{f}$ of the RMSE (Root Mean Squared Error). Box plots are made with the 50 percentage values. Upper left: $\nu=5/2$. Upper right: $\nu=3/2$. Bottom: $\nu=1/2$}}
\centering
\vspace{0.5cm}
\includegraphics[scale=0.35]{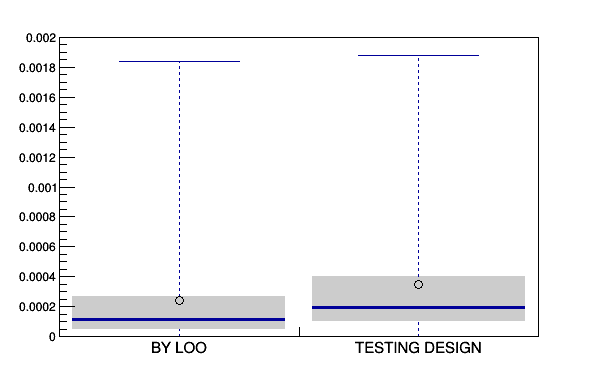}
\includegraphics[scale=0.35]{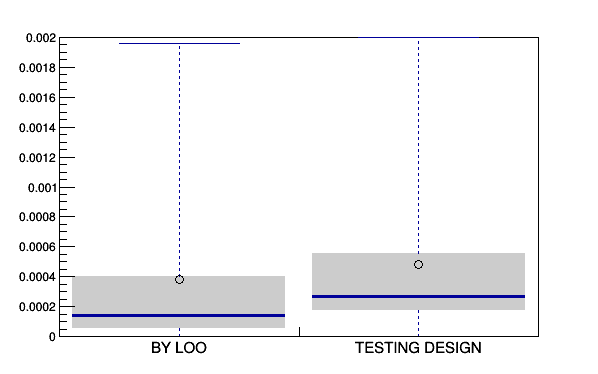}
\includegraphics[scale=0.35]{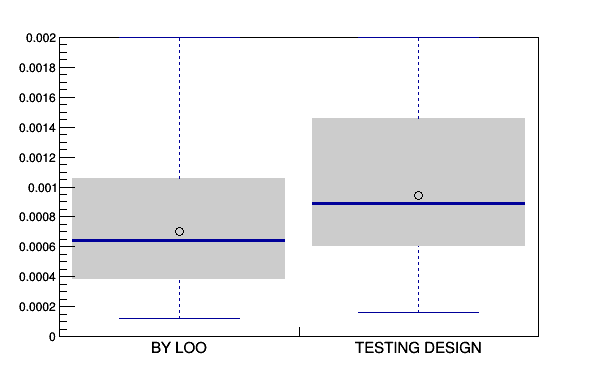}
\label{boxplots_rmse}
\end{figure}

\medskip
Every GP emulator at $\mb{x}_i^{f}$ ($1\leq i\leq 50$) was constructed with a constant mean (\ref{constantmean}) and the Mat\'ern covariance function (\ref{MaternMult}). Several values of $\nu$ were tested comprising $\nu=5/2$, $\nu=3/2$ and $\nu=1/2$. The \textit{Modeler} module of the URANIE platform was used to estimate the vector $\theta_i$ of hyperparameters by maximizing the likelihood with respect to the learning simulations (see \ref{GPappendix}). Then, both the predictive mean and variance in Equation (\ref{gp_pred}) can be evaluated at any $\lambda\in\mathcal{L}$.
The goodness-of-fit of every GP was assessed through the predictive coefficient $Q^{2}$ which measures the part of variance of the true code outputs explained by the predictive means, and also through the RMSE (Root Mean Squared Error) as a measure of bias in the predictions$:$
\begin{equation}
\scriptsize{Q_{test}^{2}(\mb{x}_i^f)=1-\frac{\sum_{s=1}^{M}\big(Y_{\mb{x}_i^{f}}(\lambda^{test}_s)-\mu^{M}_{\mb{x}_i^{f},\hat{\theta}_i}(\lambda^{test}_s)\big)^{2}}{\sum_{s=1}^{M}\big(Y_{\mb{x}_i^{f}}(\lambda^{test}_s)-\bar{Y}^{test}_{\mb{x}_i^{f}}\big)^{2}}
\,\,\,\,\,\,\,\,\,
RMSE_{test}(\mb{x}_i^f)=\sqrt{\frac{1}{M}\sum_{s=1}^{M}\big(Y_{\mb{x}_i^{f}}(\lambda^{test}_s)-\mu^{M}_{\mb{x}_i^{f},\hat{\theta}_i}(\lambda^{test}_s)\big)^{2}}}
\end{equation}
with 
$\bar{Y}^{test}_{\mb{x}_i^{f}}$ being the mean of the $M$ simulations run over the testing design $\mb{D}^{test}_M$. 
Same formula can be applied to $\mb{D}_{M}$ using leave-one-out (loo) predictions$:$
\begin{equation}
\scriptsize{Q_{loo}^{2}(\mb{x}_i^f)=1-\frac{\sum_{s=1}^{M}\big(Y_{\mb{x}_i^{f}}(\lambda_s)-\mu^{M-s}_{\mb{x}_i^{f},\hat{\theta}_i}(\lambda_s)\big)^{2}}{\sum_{s=1}^{M}\big(Y_{\mb{x}_i^{f}}(\lambda_s)-\bar{Y}_{\mb{x}_i^{f}}\big)^{2}}
\,\,\,\,\,\,\,\,\,
RMSE_{loo}(\mb{x}_i^f)=\sqrt{\frac{1}{M}\sum_{s=1}^{M}\big(Y_{\mb{x}_i^{f}}(\lambda_s)-\mu^{M-s}_{\mb{x}_i^{f},\hat{\theta}_i}(\lambda_s)\big)^{2}}}
\end{equation}
where
\begin{itemize}
\item $\mu^{M-s}_{\mb{x}_i^{f},\hat{\theta}_i}(.)$ is the predictive mean computed with respect to the $M$ learning simulations except the $s$-th,
\item $\bar{Y}_{\mb{x}_i^{f}}$ is the mean of the $M$ simulations run over $\mb{D}_M$.
\end{itemize}
The closer the $Q^{2}$ values are to $1$, the better the GP predictive means are. Almost all the values of $Q_{loo}^{2}(\mb{x}_i^f)$ and $Q_{test}^{2}(\mb{x}_i^f)$ for $1\leq i\leq 50$ are above $0.99$ regardless of the value of $\nu$. Then, Figure \ref{boxplots_rmse} presents the values of $RMSE_{loo}(\mb{x}_i^f)$ and $RMSE_{test}(\mb{x}_i^f)$. We can see that the values corresponding to $\nu=1/2$ are twice as large as those corresponding to $\nu=3/2$ and $\nu=5/2$. The latter $\nu$ are thus the most promising for point predictions.

\begin{figure}[ht!]
\caption{\textit{Distribution over $\mb{X}^{f}$ of the $95$\% coverage measure. Box plots are made with the 50 percentage values. Upper left: $\nu=5/2$. Upper right: $\nu=3/2$. Bottom: $\nu=1/2$}}
\centering
\vspace{0.5cm}
\includegraphics[scale=0.35]{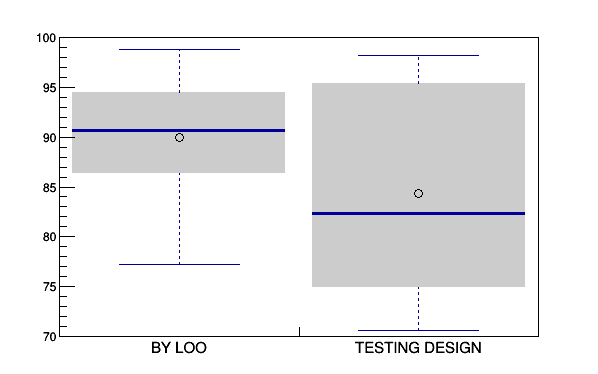}
\includegraphics[scale=0.35]{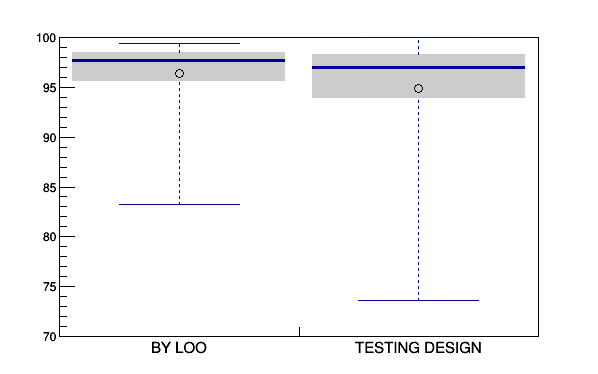}
\includegraphics[scale=0.35]{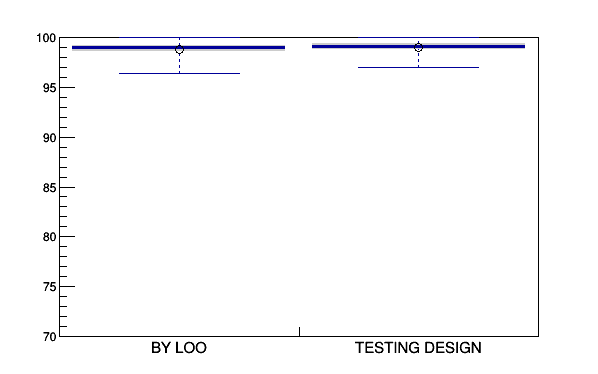}
\label{boxplots}
\end{figure}

In a second stage, the standardized prediction errors (s.p.e.) were computed to evaluate the reliability of the covariance function.
The value of $\nu$ should be judged as acceptable either if the empirical distribution of the s.p.e. matches that of the standard Gaussian distribution $\Phi$ or if most of the s.p.e. are closer to $0$ than expected. The latter is when the predictive variances have been overestimated.
For each GP emulator at $\mb{x}_i^f\in\mb{X}^f$, we calculated a coverage measure giving the percentage of the s.p.e. covered by the $95\%$ bilateral interval of $\Phi$.
Figure \ref{boxplots} displays box plots of these percentages over $\mb{X}^{f}$. Overall, the best agreement with $95\%$ is achieved with $\nu=3/2$ whereas $\nu=5/2$ and $\nu=1/2$ show respectively under-coverage and over-coverage. We can conclude that the predictive variances of the GP emulators fitted with $\nu=5/2$ are underestimated (under conservative predictions) and those fitted with $\nu=1/2$ are overestimated (conservative predictions).  

\medskip
Several GP-based MH within Gibbs samplers  (see Section \ref{The_GP-based_MH}) were run to assess the possible impact of $\nu$ on the shape of the resulting GP-based posterior distribution. We considered three settings$:$
\begin{enumerate}[1.]
\item \underline{Setting $1$:} The GP-based MH within Gibbs algorithm relies on GP emulators constructed with the value $\nu=5/2$ for every $\mb{x}_i^{f}\in\mb{X}^{f}$;
\item \underline{Setting $2$:} The GP-based MH within Gibbs algorithm relies on GP emulators constructed with the value $\nu=3/2$ for every $\mb{x}_i^{f}\in\mb{X}^{f}$;
\item \underline{Setting $3$:} similar to Setting 2, but the value $\nu=1/2$ is used for sites $\mb{x}_i^f\in\mb{X}^{f}$ where the coverage measure with $\nu=3/2$ is below $95\%$. This is a conservative option to avoid any underestimated predictive variances.
\end{enumerate}

The Gaussian-inverse gamma prior (\ref{inv_gamm_prior}) was used with $\epsilon=10^{-2}$ like in the linear case. For each setting above, $3\times 10^{6}$ MCMC samples were generated by the GP-based MH within Gibbs algorithm.
As the autocorrelation was stronger than in the linear case, illustrated by a poor mixing of the samples, four chains were run in parallel from different starting locations. In Setting $1$, only the second chain passed the Geweke testing at level $5\%$. In Setting $2$, the third and fourth chains passed it. In Setting $3$, all the chains passed it except the fourth one. The Gelman and Rubin's statistic was also computed to ensure that the four chains actually converged to the posterior distribution that they have in common, as it should in theory (see \ref{MCMCconv}).

\medskip
Among the chains that passed the Geweke testing, we selected in each setting the one having the highest e.s.s. after cutting off the burn-in period. Tables \ref{result_bayes_cosi_nonlinear_means} and \ref{result_bayes_cosi_nonlinear_vars} present respectively the posterior means of $m_1$, $m_2$ and $\sigma_1^2$, $\sigma_2^2$ along with the corresponding confidence intervals at $95\%$. We have expected that both $\sigma_1^2$ and $\sigma_2^2$ were higher in Settings $2$ and $3$ than in Setting $1$ because the predictive variances of the GP emulators are larger in those two settings. Actually, Table \ref{result_bayes_cosi_nonlinear_vars} shows that only the posterior mean of $\sigma^2_2$ is increased. Table \ref{result_bayes_cosi_nonlinear_fluctuation} presents the resulting fluctuation intervals computed in each setting. The fluctuation interval of $\Lambda_2$ is then much more spread out in both Settings $2$ and $3$ than in Setting $1$ whereas that of $\Lambda_1$ does not change much (and even a little bit narrower). A preliminary sensitivity analysis of the COSI database showed that $\Lambda_2$ is applied to the condensation model that most impacts the total condensation flow rate. This factor is thus more affected by changes into the GP-based posterior distribution (\ref{cond_latent_pg}).

Since the GP emulators in Setting $1$ yield underestimated predictive variances, the fluctuation intervals of Setting $2$ and, even more, Setting $3$ should be preferred over that of Setting $1$. This is even more critical within the BEPU framework where conservative fluctuation intervals are wanted.

\medskip
The predictive variances of the GP emulators in Settings $2$ and $3$ are on average about equal to $10^{-7}$ while the variance of $z_i^f$ is equal to $\sigma_{\epsilon_i}^2=10^{-8}$ ($1\leq i\leq 50$). Therefore, a significant gap may still exist between the fluctuation intervals reported in Table \ref{result_bayes_cosi_nonlinear_fluctuation} and those that would have been computed from the exact posterior distribution if the CATHARE 2 simulations had been instantaneous (or at least much faster) (see Remark \ref{closenessGP_exact}). 




%


\begin{table}
\caption{Posterior mean estimates $\hat{m}_1$ and $\hat{m}_2$, the corresponding confidence intervals and also the median estimates $\exp{\hat{m}_1}$ and $\exp{\hat{m}_2}$ of $\Lambda_1$ and $\Lambda_2$}
\vspace{0.5cm}
\renewcommand{\arraystretch}{0.92}
\small
\centering
\begin{tabular}{|l|l|l|c|c|c|c|}
\hline
 & $\hat{m}_1$ & $IC_{95\%}(\ee[m_1|\mb{z}^{f}])$ & $\hat{m}_2$ & $IC_{95\%}(\ee[m_2|\mb{z}^{f}]$) & $\exp{\hat{m}_1}$ & $\exp{\hat{m}_2}$ \\
\hline\hline
Setting 1 & $0.7602$ & $[0.7518,0.7686]$ & $-1.5071$ & $[-1.5353,-1.4789]$ & $2.1387$ & $0.2216$ \\
\hline
Setting 2 & $0.6963$ & $[0.6863,0.7064]$  & $-1.3814$  & $[-1.4145,-1.3484]$ & $2.0063$ & $0.2512$ \\
\hline
Setting 3 & $0.6959$  & $[0.6876,0.7043]$  & $-1.3706$  & $[-1.4003,-1.3408]$ & $2.0055$ & $0.2540$ \\
\hline
\end{tabular}
\label{result_bayes_cosi_nonlinear_means}
\end{table}

\begin{table}
\caption{Posterior mean estimates of $\sigma_1^2$ and $\sigma_2^2$ and their corresponding confidence interval}
\vspace{0.5cm}
\renewcommand{\arraystretch}{0.92}
\small
\centering
\begin{tabular}{|l|l|c|c|c|}
\hline
 & $\hat{\sigma}_1^2$ & $IC_{95\%}(\ee[\sigma_1^2|\mb{z}^{f}])$ & $\hat{\sigma}_2^2$ & $IC_{95\%}(\ee[\sigma_2^2|\mb{z}^{f}]$)  \\
\hline\hline
Setting 1 & $0.1592$ & $[0.1564,0.1621]$ & $0.8883$ & $[0.8564,0.9203]$ \\
\hline
Setting 2 & $0.1528$ & $[0.1463,0.1593]$  & $1.4785$  & $[1.4305,1.5265]$ \\
\hline
Setting 3 & $0.1474$ & $[0.1420,0.1528]$  & $1.5365$  & $[1.4877,1.5852]$ \\
\hline
\end{tabular}
\label{result_bayes_cosi_nonlinear_vars}
\end{table}

\begin{table}
\caption{$95\%$-fluctuation intervals of $\Lambda_1$ and $\Lambda_2$}
\vspace{0.5cm}
\renewcommand{\arraystretch}{0.92}
\small
\centering
\begin{tabular}{|l|c|c|}
\hline
 & $IF_{95\%}(\Lambda_1)$ & $IF_{95\%}(\Lambda_2)$  \\
\hline\hline
Setting 1 &  $[0.82,4.80]$ & $[0.02,1.59]$ \\
\hline
Setting 2 & $[0.73,4.44]$ & $[0.01,2.93]$ \\
\hline
Setting 3 & $[0.76,4.39]$ & $[0.01,3.03]$  \\
\hline
\end{tabular}
\label{result_bayes_cosi_nonlinear_fluctuation}
\end{table}



\section{Conclusions}

\label{concl_sec}

\subsection{Summary}

\medskip
The paper has laid out a Bayesian methodology to quantify the uncertainty of the physical models which close the balance equations of thermal-hydraulic system codes. This work should be received as the Bayesian implementation of the CIRCE method in that the joint posterior probability distribution of $(m,\sigma^2)$ is computed instead of a point estimate $(\hat{m},\hat{\sigma}^2)$. The interest in doing so is that the statistical uncertainty of $(\hat{m},\hat{\sigma}^2)$ can be taken into account in computing the fluctuation intervals of the multiplicative factors applied to the physical models.
The fluctuation intervals are then expected to be more spread out than those derived from the standard CIRCE method, which is particularly relevant for BEPU studies where some degree of conservatism is required in assessing uncertainties.
In addition, we have tackled the extended Bayesian framework where the linear assumption of the code outputs with respect to the factors has been dropped.

\medskip
An efficient MCMC algorithm, known as the blocked Gibbs sampler, has been implemented both in the linear and non linear settings to generate samples of the posterior distribution of $(m,\sigma^2)$. As the cost of this algorithm strongly increases with the time duration of the thermal-hydraulic simulations, a Gaussian process(GP)-based blocked Gibbs sampler has been implemented in the non linear setting. Unlike other surrogate models such as neural networks, polynomial chaos expansion, the GP emulator is particularly suited for computer experiments because it can interpolate the learning simulations as well as providing Gaussian predictions of the code outputs at any other realization of the multiplicative factors. The GP-based blocked Gibbs sampler relies on a conditional posterior distribution of the missing (log)-values of the multiplicative factors in which the predictive means and variances of the GP emulators replace the expensive simulations. This sampler has been applied to the COSI tests to quantify the uncertainty of two condensation models at a safety injection.

A great attention has been paid on the choice of the covariance function so that the actual smoothness of the code outputs is captured as best as possible. Several Mat\'ern covariance functions have been tested on the COSI simulations. The smaller the smoothness parameter $\nu$, the less smooth the Gaussian process trajectories. The values $\nu=1/2$ and $\nu=3/2$ have yielded much more robust predictions of the condensation flow rate than the value $\nu=5/2$ for which the predictive variances have been underestimated. As a consequence, the fluctuation intervals of the multiplicative factors computed with $\nu=5/2$ (Setting $1$) have been different than those computed with a more suited $\nu=3/2$ (Setting $2$) or even a more conservative option (Setting $3$).

\medskip
The more the GP predictions match the true values of the code outputs, the more the GP-based posterior distribution is expected to be close to the exact posterior distribution (\ref{post_expression_nonlinear}) in some sense, for example, according to the Kullback-Leibler divergence. We could surely prove such a convergence property in the same way than done in \citet{Damblin18}. The proof would rely on the theory of kernel interpolation \citep{Schaback95}.
From a practical point of view, however, when the number $M$ of learning simulations is moderate, it remains challenging to guaranty that the actual gap between both of those distributions is negligible or even not that large. 




\subsection{Future works} 

According to the previous discussion, the improvement of the accuracy of the GP emulators should be aimed for. This can simply be done by increasing the number of learning simulations so that the predictive variances are significantly down in comparison with the variance of the experimental error. However, the computational cost of the GP-based Gibbs algorithm would increase (see \ref{GP_computation_cost}). We could suggest two ways for facing this limitation$:$ evaluating the GP emulators in parallel at every iteration of the Gibbs sampler and speeding up GP evaluations for large datasets \citep{Gram15,Rulliere17}. 
 
 
Another tough question concerns the choice of the prior distribution for the variances $\sigma^2$ of the multiplicative factors. If few informative inverse-gamma priors are put on such parameters, as we did in the paper, then the significant prior mass near $0$ does not vanish in the posterior distribution. This can lead to a Bayesian procedure that is not as much as data-dominated as expected, especially in small data situations. Some possibly more suited prior distributions are promoted in \citep{Gelman06} and should be tested.

Finally, because the posterior samples generated by the GP-based blocked Gibbs algorithm are highly correlated, other proposal densities for the Metropolis-Hastings algorithm could be tested or even advanced samplers such a Delayed Rejection Adaptive Metropolis (DRAM) could be implemented \citep{Haario06}. In this way, we should improve the mixing speed of the Markov chain and, therefore, the precision of point estimates as well as resulting fluctuation intervals of the multiplicative factors.

\section{Acknowledgments}

This work has been funded by the tripartite project devoted to Uncertainty Quantification, consisting of French Alternative Energies and Atomic Energy Commission (CEA), Electricity of France (EDF) and Framatome (FRA).
The authors thank the URANIE team for guidance in using the software and careful reading of the paper, as well as Lucia Sargentini and Alberto Ghione for fruitful discussions on the COSI tests. Finally, we would like to thank the two anonymous reviewers who help improving the content of the paper significantly.

\section*{References}

\bibliographystyle{elsarticle-harv}

\bibliography{biblio_article_circe}

\newpage

\appendix

\appendixpage

\section{(Linear case) The different MCMC samplers possible}

In this first appendix, we explain in details how the posterior distribution of the parameters can be computed in the linear case (Section \ref{Bayes_lin_sec} of the main document). We have identified at least three ways$:$
\begin{itemize}
\item run a regular Gibbs sampler ,
\item run a blocked Gibbs sampler,
\item after integrating the complete likelihood out the missing data $\mb{A}^{\prime}$, compute the conditional posterior distribution of $b$ with respect to $\sigma^2$, then run an MCMC algorithm such as MH algorithm or Delayed Rejection Adaptive Metropolis (DRAM), to generate samples of the marginal posterior distribution of $\sigma^2$.
\end{itemize}

\label{MCMC_several_lin}

\medskip
\vspace{3pt}\hrule\vspace{6pt}
\noindent \textbf{Regular Gibbs sampler} 
\vspace{3pt}\hrule\vspace{6pt}

\begin{enumerate}[1.]
\item Set starting values: $(b(0),\sigma^2(0))$
\item Choose a number $N_{mcmc}$ of iterations, then for $1\leq k\leq N_{mcmc}$ generate in a loop$:$
\begin{equation}
\label{fcd1}
\mb{A}^{\prime}(k+1) \thicksim \mb{A}^{\prime} \,|\, b(k), \sigma^2(k),\mb{z}^{\prime,f}, \mb{X}^f
\end{equation}
\begin{equation}
\label{fcd2}
b(k+1)\thicksim b \,|\, \sigma^2(k), \mb{A}^{\prime}(k+1), \mb{z}^{\prime,f}, \mb{X}^f
\end{equation}
\begin{equation}
\label{fcd3}
\sigma^2(k+1)\thicksim \sigma^2 \,|\, b(k+1), \mb{A}^{\prime}(k+1),\mb{z}^{\prime,f}, \mb{X}^f
\end{equation}
\end{enumerate}
\vspace{3pt}\hrule\vspace{6pt}

\bigskip
Equations (\ref{fcd1}), (\ref{fcd2}) and (\ref{fcd3}) are the full conditional posterior distributions of $\mb{A}^{\prime}$, $b$ and $\sigma^2$ respectively.
This algorithm theoretically converges to $b,\sigma^2,\mb{A}^{\prime}|\mb{z}^{\prime,f},\mb{X}^{f}$. 


\medskip
\begin{thm}
\label{proofGibbs}
Using the Gaussian-inverse gamma prior distribution (\ref{inv_gamm_prior}), the full conditional distributions (\ref{fcd2}), (\ref{fcd3}) and (\ref{fcd1}) are equal to respectively$:$
\begin{equation}
\label{cond1_linear}
b|\sigma^2,\mb{A}^{\prime},\mb{z}^{\prime,f},\mb{X}^{f} \thicksim \otimes_{j=1}^{p} \mathcal{N}\big(\frac{a_j}{a_j+n}\mu_j+\frac{n}{a_j+n}\bar{\alpha}_j^{\prime},\frac{\sigma_{j}^{2}}{a_j+n}\big)\,\,\,\textrm{with}\,\,\,
\bar{\alpha}_j^{\prime}=\frac{1}{n}\sum_{i=1}^{n} \alpha_{i,j}^{\prime}
\end{equation}

\begin{equation}
\label{cond2_linear}
\sigma^{2}|b,\mb{A}^{\prime},\mb{z}^{\prime,f},\mb{X}^{f}\thicksim \otimes_{j=1}^{p} \mathcal{IG}(\psi_j+\frac{n}{2}+\frac{1}{2}, \frac{1}{2}\sum_{i=1}^{n} (\alpha^{\prime}_{i,j}-b_{j})^2+\frac{a_j}{2}(b_{j}-\mu_j)^2+\gamma_j)
\end{equation}

\begin{equation}
\label{cond3_linear}
\mb{A}^{\prime}|b,\sigma^2,\mb{z}^{\prime,f},\mb{X}^{f} \thicksim \otimes_{i=1}^{n} \mathcal{N}(S_i\nu_i,S_i)
\end{equation}
where $$S_i^{-1}=\frac{h(\mb{x}_i^f)h^{T}(\mb{x}_i^f)}{\sigma^2_{\epsilon_i}}+\Sigma^{-1}\,\,\,\,\in\rr^{p\times p}\,\,\,\,\,\textrm{with}\,\,\,\,\Sigma=\text{diag}(\sigma^2)=\text{diag}(\sigma_1^2,\cdots,\sigma_p^2)$$ and $$\nu_i=\frac{h(\mb{x}_i^f)z_i^{\prime,f}}{\sigma^2_{\epsilon_i}}+\Sigma^{-1}b\,\,\,\,\in\rr^{p}.$$ 
\end{thm}

\begin{pf}

The posterior distribution can be expanded as
\begin{equation}
\label{full_post}
\pi(b,\sigma^{2},\mb{A}^{\prime}|\mb{z}^{\prime,f},\mb{X}^{f})\propto \pi(\mb{z}^{\prime,f}|\mb{A}^{\prime},\mb{X}^{f},b,\sigma^{2})\pi(\mb{A}^{\prime}|b,\sigma^{2})\pi(b,\sigma^{2}) 
\end{equation}
where
\begin{align}
\label{cond_field}
\pi(\mb{z}^{\prime,f}|\mb{A}^{\prime},\mb{X}^f,b,\sigma^{2})=&\prod_{i=1}^{n}\Big(\frac{1}{\sqrt{2\pi}\sigma_{\epsilon_i}}\Big)\exp{\Big\{-\frac{1}{2}\sum_{i=1}^{n} \frac{(z_i^f-h^{T}(\mb{x}_i^f)\alpha^{\prime}_i)^{2}}{\sigma^2_{\epsilon_i}}\Big\}}
\end{align}
\begin{align}
\nonumber
\pi(\mb{A}^{\prime}|b,\sigma^{2})=& \Big(\frac{|\Sigma^{}|^{-1/2}}{(2\pi)^{p/2}}\Big)^{n}\exp{\Big\{-\frac{1}{2}\sum_{i=1}^{n}(\alpha^{\prime}_i-b)^{T}\Sigma^{-1}(\alpha^{\prime}_i-b)\Big\}} \\
=&  \frac{\prod_{j=1}^{p}(\sigma_{j}^{})^{-n}}{(2\pi)^{pn/2}} \exp{\Big\{-\frac{1}{2}\sum_{i=1}^{n} \sum_{j=1}^{p} \frac{(\alpha_{i,j}^{\prime}-b_{j})^2}{\sigma_{j}^{2}}\Big\}}
\label{cond_lambda}
\end{align}
and
\begin{align}
\nonumber
\pi(b|\sigma^{2})=& \prod_{j=1}^{p} \pi(b_{j}|\sigma_{j}^{2})  \\
=& \frac{1}{(2\pi)^{p/2}}\prod_{j=1}^{p} \frac{\sqrt{a_j}}{\sigma_{j}}\times\exp{\Big\{-\frac{1}{2}\sum_{j=1}^{p}\frac{a_j(b_{j}-\mu_j)^{2}}{\sigma_{j}^{2}}\Big\}}
\label{priorM}
\end{align}

\begin{equation}
\pi(\sigma^{2})= \prod_{j=1}^{p} \pi(\sigma_{j}^{2})
\propto\prod_{j=1}^{p} \frac{(\sigma_{j})^{2(-\psi_j-1)}}{\Gamma(\psi_j)}\times\exp{\Big\{\sum_{j=1}^{p}-\frac{\gamma_j}{\sigma_{j}^{2}}\Big\}}
\label{priorS2}
\end{equation}
By gathering together Equations (\ref{cond_lambda}), (\ref{priorM}) et (\ref{priorS2}) where $\sigma^{2}$ occurs, we have
\begin{multline}
\pi(\sigma^{2}|b,\mb{A}^{\prime},\mb{z}^{\prime,f},\mb{X}^{f})\propto \prod_{j=1}^{p} (\sigma_{j})^{-\big(n+1+2(\psi_j+1)\big)}
\,\,\times \\
\exp{\Big\{-\sum_{j=1}^{p} \frac{\sum_{i=1}^{n}(\alpha_{i,j}^{\prime}-b_{j})^{2}}{2\sigma^{2}_{j}}+\frac{a_j(b_{j}-\mu_{j})^{2}}{2\sigma_{j}^{2}}+\frac{\gamma_j}{\sigma_{j}^{2}}\Big\}}
\end{multline}
We can recognize the probability density of  $p$ independent Gaussian-inverse gamma distributions with parameters 
\begin{equation}
\psi_j+\frac{n}{2}+\frac{1}{2}
\end{equation}
and 
\begin{equation}
\frac{1}{2}\sum_{i=1}^{n} (\alpha^{\prime}_{i,j}-b_{j})^2+\frac{a_j}{2}(b_{j}-\mu_j)^2+\gamma_j,
\end{equation}
which completes the proof of (\ref{cond2_linear}).
Now, by gathering together Equations (\ref{cond_lambda}) and (\ref{priorM}) containing $b$, we have
\begin{align}
\pi(b|\sigma^{2},\mb{A}^{\prime},\mb{z}^{\prime,f},\mb{X}^{f})\propto & \exp{\Big\{-\sum_{j=1}^{p} \frac{\sum_{i=1}^{n}(\alpha_{i,j}^{\prime}-b_{j})^{2}}{2\sigma_{j}^{2}}+\frac{a_j(b_{j}-\mu_{j})^{2}}{2\sigma_{j}^{2}}\Big\}} \\
\propto & \exp{\Big\{-\sum_{j=1}^{p}\frac{\sum_{i=1}^{n}
\alpha_{i,j}^{\prime^2}-2b_{j}\sum_{i=1}^{n}\alpha_{i,j}^{\prime}+nb^{2}_{j}+a_jb^{2}_{j}-2a_jb_{j}\mu_j+a_j\mu_j^2}{2\sigma_{j}^{2}} \Big\}} \\
\propto & \exp{\Big\{-\sum_{j=1}^{p}\frac{(a_j+n)}{2\sigma_{j}^{2}} \Big[b^{2}_{j}-\frac{2b_{j}(n\bar{\alpha}_j^{\prime}+a_j\mu_j)}{a_j+n}+\frac{\sum_{i=1}^{n}
\alpha_{i,j}^{\prime^2}+a_j\mu_j^2}{a_j+n}\Big]\Big\}} \\
\propto & \exp{\Big\{-\sum_{j=1}^{p}\frac{(a_j+n)}{2\sigma_{j}^{2}} \Big[b_{j}-\frac{n\bar{\alpha}_j^{\prime}+a_j\mu_j}{a_j+n}\Big]^2\Big\}}
\end{align}
which proves Equation (\ref{cond1_linear}). Lastly, proof of (\ref{cond3_linear}) is based on combining Equation (\ref{cond_field}) with Equation (\ref{cond_lambda})
\begin{equation}
\label{eq1_appendix}
\pi(\mb{A}^{\prime}|b,\sigma^{2},\mb{z}^{\prime,f},\mb{X}^{f}) \propto \exp{\Big\{-\frac{1}{2}\sum_{i=1}^{n}  \frac{(z_i^{\prime,f}-h^{T}(\mb{x}_i^f)\alpha^{\prime}_i)^{2}}{\sigma_{\epsilon_i}^2}-
\frac{1}{2}(\alpha^{\prime}_i-b)^{T}\Sigma^{-1}(\alpha^{\prime}_i-b)\Big\}} 
\end{equation}
Equation (\ref{eq1_appendix}) is rewritten as
\begin{multline}
\label{cond_field_dev}
\pi(\mb{A}^{\prime}|b,\sigma^{2},\mb{z}^{\prime,f},\mb{X}^{f}) \propto \exp{\Big\{\sum_{i=1}^{n}\Big(-\frac{1}{2}\alpha^{\prime^T}_i\Big[ \frac{h(\mb{x}_i^f)h^{T}(\mb{x}_i^f)}{\sigma^2_{\epsilon_i}}+\Sigma^{-1}\Big]\alpha^{\prime}_i}+ \\
\Big[\frac{h(\mb{x}_i^f)z_i^{\prime,f}}{\sigma_{\epsilon_i}^2}+\Sigma^{-1}b\Big]^{T}\alpha^{\prime}_i\Big)\Big\}
\end{multline}
Then, if for any $\nu\in\rr^{n}$ and any $B\in\rr^{n\times n}$ positive definite matrix
\begin{equation}
\label{lemma}
f(w)\propto \exp{\Big\{-\frac{1}{2} w^{T}B^{-1}w+\nu^{T}w\Big\}}
\end{equation}
then $W\thicksim\mathcal{N}(B\nu,B)$.
Proof of (\ref{cond3_linear}) is done by applying this lemma to (\ref{cond_field_dev}).

\end{pf}
The posterior distribution of $(m,\sigma^2)$ is derived from that of $(b,\sigma^2)$ by shifting of $\alpha^{\star}$ the posterior samples $\{b(k)\}_k$.

\medskip
A slightly different algorithm can be performed instead of the regular Gibbs sampler. It is referred to as the blocked Gibbs sampler$:$

\medskip
\vspace{3pt}\hrule\vspace{6pt}
\noindent \textbf{Blocked Gibbs sampler} 
\vspace{3pt}\hrule\vspace{6pt}

\begin{enumerate}[1.]
\item Set starting values: $(b(0),\sigma^2(0))$
\item Choose a number $N_{mcmc}$ of iterations. Then, for $1\leq k\leq N_{mcmc}$, generate in a loop:
\begin{equation}
\label{fcd1_blocked}
\mb{A}^{\prime}(k+1) \thicksim \mb{A}^{\prime}\,|\, b(k),\sigma^2(k),\mb{z}^{\prime,f}, \mb{X}^f
\end{equation}
\begin{equation}
\label{fcd2_blocked}
b(k+1),\sigma^2(k+1)\thicksim b,\sigma^2 \,|\, \mb{A}^{\prime}(k+1),\mb{z}^{\prime,f}, \mb{X}^f
\end{equation}
\end{enumerate}
\vspace{3pt}\hrule\vspace{6pt}

\begin{thm}
The parameters $(b,\sigma^2)$ are conditionally conjugated with respect to the unobserved samples $\mb{A}^{\prime}$, meaning that the conditional distribution $b,\sigma^2|\mb{A}^{\prime},\mb{z}^{\prime,f},\mb{X}^f$ follows a Gaussian-inverse gamma distribution whose density is such that
\begin{equation}
\label{cond1IGN}
\pi(b,\sigma^2|\mb{A}^{\prime},\mb{z}^{\prime,f},\mb{X}^f)=\pi(b|\sigma^2,\mb{A}^{\prime},\mb{z}^{\prime,f},\mb{X}^f)\pi(\sigma^2|\mb{A}^{\prime},\mb{z}^{\prime,f},\mb{X}^f)
\end{equation}
with $b|\sigma^2,\mb{A}^{\prime},\mb{z}^{\prime,f},\mb{X}^f$ being a Gaussian distribution with parameters given by Equation (\ref{cond1_linear}) and
\begin{equation}
\label{cond2IGN}
\sigma^{2}|\mb{A}^{\prime},\mb{z}^{\prime,f},\mb{X}^{f}\thicksim \otimes_{j=1}^{p}\mathcal{IG}\Big(\psi_j+\frac{n}{2},\gamma_j+\frac{1}{2}\Big(\sum_{i=1}^{n}(\alpha_{i,j}^{\prime}-\bar{\alpha}_j^{\prime})^{2}+ \frac{a_j n (\bar{\alpha}_j^{\prime}-\mu_j)^{2}}{a_j+n}\Big)\Big)
\end{equation}
\end{thm}
\begin{pf}
This result derives from the conjugacy of the Gaussian-inverse gamma distribution (see for instance \citet{Murphy07} for details concerning conjugate Bayesian analyses). First, we need to rewrite Equation (\ref{cond_lambda}) by introducing $\bar{\alpha}_j^{\prime}$ in it, which gives
\begin{align}
\nonumber
\pi(\mb{A}^{\prime}|b,\sigma^{2})
\propto &  \prod_{j=1}^{p}(\sigma_{j}^{})^{-n} \exp{\Big\{-\frac{1}{2}\sum_{i=1}^{n} \sum_{j=1}^{p} \frac{(\alpha_{i,j}^{\prime}-b_{j})^2}{\sigma_{j}^{2}}\Big\}}\\
\nonumber
\propto &  \prod_{j=1}^{p}(\sigma_{j}^{})^{-n}\exp{\Big\{-\frac{1}{2}\sum_{j=1}^{p}} \sum_{i=1}^{n} \frac{(\alpha_{i,j}^{\prime}-\bar{\alpha}_j^{\prime}+\bar{\alpha}_j^{\prime}-b_{j})^2}{\sigma_{j}^{2}}\Big\}\\
\nonumber
\propto &  \prod_{j=1}^{p}(\sigma_{j}^{})^{-n}\exp{\Big\{-\frac{1}{2}\sum_{j=1}^{p}} \sum_{i=1}^{n} \frac{(\alpha_{i,j}^{\prime}-\bar{\alpha}_j^{\prime})^2+(\bar{\alpha}_j^{\prime}-b_{j})^2}{\sigma_{j}^{2}}\Big\}\\
\propto & \prod_{j=1}^{p}(\sigma_{j}^{})^{-n}\exp{\Big\{-\frac{1}{2}\sum_{j=1}^{p}}\frac{ns_j+n(\bar{\alpha}_j^{\prime}-b_j)^{2}}{\sigma_{j}^{2}}\Big\}
\label{cond_lambda_here}
\end{align}
where $s_j=\frac{1}{n}\sum_{i=1}^{n} (\alpha_{i,j}^{\prime}-\bar{\alpha}_j^{\prime})^2$. Then, the density of the joint posterior distribution of $b$ and $\sigma^2$ conditional on $\mb{A}^{\prime}$ is equal to the product between Equations (\ref{cond_lambda_here}), (\ref{priorM}) et (\ref{priorS2})$:$
\begin{multline}
\label{post_joint_b_sig2}
\pi(b,\sigma^2|\mb{A}^{\prime},\mb{z}^{\prime,f},\mb{X}^f)\propto \prod_{j=1}^{p}(\sigma_j)^{-n-1-2(\psi_j+1)}
\exp{\Big\{-\frac{1}{2\sigma_j^2}\sum_{j=1}^{p}\Big[
n(\bar{\alpha}_j^{\prime}-b_j)^{2}+a_j(b_j-\mu_j)^2\Big]\Big\}} \times \\
\exp{\Big\{-\sum_{j=1}^{p}\frac{\gamma_j}{\sigma_j^2}+\frac{ns_j}{2\sigma_j^2}\Big\}}
\end{multline}
The terms between brackets in the first sum can be expanded, then factorized in the following way$:$
\begin{align}
n(\bar{\alpha}_j^{\prime}-b_j)^{2}+a_j(b_j-\mu_j)^2
=&(a_j+n)\Big(b_j-\frac{a_j\mu_j+n\bar{\alpha}_j^{\prime}}{a_j+n}\Big)^{2}-\frac{(a_j\mu_j+n\bar{\alpha}_j^{\prime})^{2}}{a_j+n}+a_j\mu^2_j+n\bar{\alpha}_j^{\prime^{2}}\\
=&(a_j+n)\Big(b_j-\frac{a_j\mu_j+n\bar{\alpha}_j^{\prime}}{a_j+n}\Big)^{2}+\frac{n a_j(\bar{\alpha}_j^{\prime}-\mu_j)^{2}}{a_j+n}
\end{align}
It follows that Equation (\ref{post_joint_b_sig2}) can be rewritten as
\begin{multline}
\label{lastBlockedProof}
\pi(b,\sigma^2|\mb{A}^{\prime},\mb{z}^{\prime,f})\propto \prod_{j=1}^{p}(\sigma_j)^{-1}(\sigma^2_j)^{-\frac{n}{2}-\psi_j-1}\times \\
\exp{\Big\{-\frac{1}{2\sigma_j^2}\sum_{j=1}^{p}\Big[
(a_j+n)\Big(b_j-\frac{a_j\mu_j+n\bar{\alpha}_j^{\prime}}{a_j+n}\Big)^{2}\Big]\Big\}} \times \\
\exp{\Big\{-\frac{1}{\sigma_j^2}\sum_{j=1}^{p}\gamma_j+\frac{ns_j}{2}+\frac{n a_j(\bar{\alpha}_j^{\prime}-\mu_j)^{2}}{2(a_j+n)}\Big\}}
\end{multline}
In Equation (\ref{lastBlockedProof}), we can recognize the product over $1\leq j\leq p$ of a Gaussian density for $b_j$ conditional on $\sigma_j^2$ multiplied by an inverse Gamma density for $\sigma_j^2$. The corresponding parameters of both these distributions are those given by Equation (\ref{cond1_linear}) and (\ref{cond2IGN}) respectively, which ends the proof.
\end{pf}

The blocked Gibbs sampler follows the hierarchical structure of Model (\ref{CIRCEmodel_lin_alpha}). According to \citet{Diebolt94}, it can converge faster to the stationary distribution than the regular Gibbs sampler. After starting both of these algorithms with values $b(0)$ and $\sigma^2(0)$ being far away from the support of the posterior distribution, we have observed on an artificial example that the blocked Gibbs sampler moved faster to the region of high posterior probability.

\medskip
The last option is to compute the posterior distribution $b,\sigma^2|\mb{X}^{f},\mb{z}^{\prime,f}$ from the likelihood of $(b,\sigma^2)$ after integrating the complete likelihood out the missing data $\mb{A}^{\prime}$. This integral given by Equation (\ref{margLikeli}) is tractable and we have
\begin{equation}
\mathcal{L}(\mb{z}^{\prime,f}|\mb{X}^{f},b,\sigma^2)=\prod_{i=1}^{n}\mathcal{L}(z_i^{\prime,f}|\mb{x}_i^{f},b,\sigma^2)
\end{equation}
with $z_i^{\prime,f}|\mb{x}_i^{f},b,\sigma^2$ being a Gaussian distribution with mean and variance equal to $h^{T}(\mb{x}_i^{f})b$ and $\sigma^2_{\epsilon_i}+h^{T}(\mb{x}_i^{f})\Sigma h(\mb{x}_i^{f})$ respectively. Then, the joint posterior distribution of $b$ and $\sigma^2$ is given by the Bayes formula$:$
\begin{equation}
\pi(b,\sigma^2|\mb{z}^{\prime,f},\mb{X}^{f})\propto \mathcal{L}(\mb{z}^{\prime,f}|\mb{X}^{f},b,\sigma^2)\pi(b,\sigma^2)
\end{equation}

\begin{thm}
\label{cond_b_sigma}
Let us define$:$
\begin{equation}
\mu:=(\mu_1,\cdots,\mu_p)^{T}
\end{equation}
\begin{equation}
\Sigma_b:=\textrm{diag}\,\Big(\frac{\sigma_1^2}{a_1},\cdots,\frac{\sigma_p^2}{a_p}\Big)
\end{equation}
\begin{equation}
\Sigma_{f,\epsilon}:=\textrm{diag}\,\Big(h^{T}(\mb{x}_1^f)\Sigma h(\mb{x}_1^f)+\sigma_{\epsilon_1}^2,\cdots,h^{T}(\mb{x}_n^f)\Sigma h(\mb{x}_n^f)+\sigma_{\epsilon_n}^2\Big)
\end{equation}
Then, the conditional posterior distribution $b|\sigma^2,\mb{z}^{\prime,f},\mb{X}^{f}$ follows a $p$-dimensional Gaussian distribution with mean denoted by the vector 
\begin{equation}
\tilde{\mu}_b=\tilde{\Sigma}_b\Big(H^{T}\Sigma^{-1}_{f,\epsilon}\mb{z}^{\prime,f}+\Sigma^{-1}_b\mu\Big)\in\rr^{p}
\end{equation}
and variance denoted by the $p\times p$ matrix
\begin{equation}
\tilde{\Sigma}_b=\Big(H^{T}\Sigma^{-1}_{f,\epsilon}H+\Sigma^{-1}_b\Big)^{-1}\in\mathcal{M}_{p,p}(\rr)
\end{equation}
\end{thm}

\begin{pf}
\begin{multline}
\label{eq1_proof_marg}
\pi(b|\sigma^2,\mb{z}^{\prime,f},\mb{X}^{f})\propto \exp{\Big\{-\frac{1}{2}(\mb{z}^{\prime,f}-Hb)^{T}\Sigma^{-1}_{f,\epsilon}(\mb{z}^{\prime,f}-Hb)\Big\}}\times \\
\exp{\Big\{-\frac{1}{2}(b-\mu)^{T}\Sigma_b^{-1}(b-\mu)\Big\}}
\end{multline}
Equation (\ref{eq1_proof_marg}) can be rewritten in the following way
\begin{equation}
\pi(b|\sigma^2,\mb{z}^{\prime,f},\mb{X}^{f})\propto \exp{\Big\{-\frac{1}{2}b^{T}\Big(H^{T}\Sigma^{-1}_{f,\epsilon}H+\Sigma^{-1}_b\Big)b+\Big[H^{T}\Sigma^{-1}_{f,\epsilon}\mb{z}^{\prime,f}+\Sigma_b^{-1}\mu\Big]^{T}b\Big\}}
\end{equation}
The lemma given by Equation (\ref{lemma}) ends the proof.
\end{pf}
To generate samples of the joint posterior distribution of $b$ and $\sigma^2$, the marginal posterior distribution of $\sigma^2$ is required. First, the marginal likelihood of $\sigma^2$ is equal to 
\begin{equation}
\mathcal{L}(\mb{z}^{\prime,f}|\mb{X}^{f},\sigma^2)=\int_{\rr^{p}}\mathcal{L}(\mb{z}^{\prime,f}|\mb{X}^{f},b,\sigma^2)\pi(b|\sigma^2)db 
\end{equation}
By keeping only the terms implying $\mb{z}^{\prime,f}$, we obtain that
\begin{multline}
\mathcal{L}(\mb{z}^{\prime,f}|\mb{X}^{f},\sigma^2) \propto  \int_{\rr^{p}} \exp{\Big\{-\frac{1}{2}(\mb{z}^{\prime,f}-Hb)^{T}\Sigma^{-1}_{f,\epsilon}(\mb{z}^{\prime,f}-Hb)\Big\}}\times \\
\exp{\Big\{-\frac{1}{2}(b-\mu)^{T}\Sigma_b^{-1}(b-\mu)db\Big\}} 
\end{multline}
The computation of this integral gives that $\mb{z}^{\prime,f}|\mb{X}^{f},\sigma^2$ is Gaussian with mean and variance equal to $H\mu$ and $\Sigma_{f,\epsilon}+H\Sigma_b H^{T}$ respectively. The marginal posterior of $\sigma^2$ is then given by the Bayes formula$:$
\begin{equation}
\pi(\sigma^2|\mb{z}^{\prime,f},\mb{X}^{f})\propto \mathcal{L}(\mb{z}^{\prime,f}|\mb{X}^{f},\sigma^2)\pi(\sigma^2)
\end{equation}
which is intractable analytically. Hence, an MCMC algorithm such as Delayed Rejection Adaptive Metropolis (DRAM) can be used to generate posterior samples of $\sigma^2$ \citep{Haario06}. Then, the corresponding posterior samples of $b$ can be generated by Theorem \ref{cond_b_sigma}. When the number $p$ of multiplicative factors $\Lambda_j$ grows up, this algorithm should converge faster to the joint posterior distribution of $(b,\sigma^2)$ than either of the previous Gibbs samplers. 
Indeed, it can generate samples from the conditional posterior distribution of $b$ with respect to $\sigma^2$, which is most efficient than sampling from the conditional posterior distribution of $b$ with respect to both $\sigma^2$ and $\mb{A}^{\prime}$. 


\section{(Non-linear case) Statistical details}

\subsection{The GP emulator}

\label{GPappendix}

As a second order stationary GP is assumed, the mean part is written as a regression model $H^{T}_i(\lambda)\beta_i$ parametrized by the vector $\beta_i\in\rr^{m}$ and the covariance function is decomposed as the product of the variance $\sigma_i^2$ and a correlation function parametrized by $l_i$$:$
\begin{equation}
K_{\mb{x}_i^f,\sigma_i^2,l_i}(\lambda,\lambda^{\prime})=\sigma_i^2 C_{\mb{x}_i^f,l_i}(\lambda,\lambda^{\prime})\,\,\,\,\,\,\,;\,\,\,\,\,\,\,\lambda,\lambda^{\prime}\in\mathcal{L}
\end{equation}
This appendix provides the mathematical expressions of both the predictive mean and variance of the posterior GP constructed at $\mb{x}_i^f\in\mb{X}^{f}$ with respect to the $M$ learning simulations $Y_{\mb{x}_i^f}(\mb{D}_M)\in\rr^{M}$. Let us denote the correlation matrix of these simulations by$:$
\begin{equation}
C_{\mb{x}_i^f,l_i}(\mb{D}_M)=\Big[C_{\mb{x}_i^f,l_i}\big(\lambda_s,\lambda_{s^{\prime}}\big)\Big]_{1\leq s,s^{\prime}\leq M}\,\,\in\,\,\mathcal{M}_{M,M}(\rr)
\end{equation}
Let $\lambda$ any input value of $\mathcal{L}$. Then, the vector of the correlations between $Y_{\mb{x}_i^f}(\lambda)$ and every component of $Y_{\mb{x}_i^f}(\mb{D}_M)$ is equal to
\begin{equation}
C_{\mb{x}_i^f,l_i}\big(\lambda,\mb{D}_M\big)=\Big(C_{\mb{x}_i^f,l_i}(\lambda,\lambda_1),\cdots,C_{\mb{x}_i^f,l_i}(\lambda,\lambda_M)\Big)^{T}\in\rr^{M}
\end{equation}
The predictive mean at $\lambda$ is written as the conditional mean of the GP emulator with respect to $Y_{\mb{x}_i^f}(\mb{D}_M)$$:$
\begin{equation}
\label{predMean}
\mu^{M}_{\mb{x}_i^f,\beta_i,l_i}(\lambda)=H^T_i(\lambda)\beta_i+C_{\mb{x}_i^f,l_i}\big(\lambda,\mb{D}_M\big)^{T}C_{\mb{x}_i^f,l_i}(\mb{D}_M)^{-1}\Big(Y_{\mb{x}_i^f}(\mb{D}_M)-m_{\mb{x}_i^f}(\lambda)\Big).
\end{equation}
Let $\lambda,\lambda^{\prime}$ be two input locations of $\mathcal{L}$. Then, the predictive covariance is given by the conditional covariance
\begin{equation}
\label{predVar}
V^{M}_{\mb{x}_i^f,l_i}(\lambda,\lambda^{\prime})=\sigma_i^{2}\Big(C_{\mb{x}_i^f,l_i}(\lambda,\lambda^{\prime})-C_{\mb{x}_i^f,l_i}\big(\lambda,\mb{D}_M\big)^{T}C_{\mb{x}_i^f,l_i}(\mb{D}_M)^{-1}C_{\mb{x}_i^f,l_i}\big(\lambda^{\prime},\mb{D}_M\big) \Big).
\end{equation}
An important property is that the GP emulator interpolates the learning simulations, meaning that for every location $\lambda_s\in\mb{D}_M$ we have for $1\leq s\leq M$
\begin{equation}
\mu^{M}_{\mb{x}_i^f,\beta_i,l_i}(\lambda_s)=Y_{\mb{x}_i^f}(\lambda_s)
\end{equation}
and
\begin{equation}
V^{M}_{\mb{x}_i^f,l_i}(\lambda_s,\lambda_s)=0.
\end{equation}
Equations (\ref{predMean}) and (\ref{predVar}) are valid when the vector of hyperparameters $\theta_i:=(\beta_i,\sigma_i^2,l_i)$ is known. In practice, however, $\theta_i$ need be estimated from the learning simulations $Y_{\mb{x}_i^{f}}(\mb{D}_M)$, most commonly by maximizing the likelihood function$:$
\begin{equation}
\hat{\theta}_i=(\hat{\beta}_{i},\hat{\sigma}^2_{i},\hat{l}_{i})=\argmax{\theta_i}{\mathcal{L}_i(Y_{\mb{x}_i^f}(\mb{D}_M)|\theta_i)}.
\end{equation}
By satisfying to a zero-gradient condition, we obtain 
\begin{equation}
\label{beta_estimate}
\hat{\beta}_i(l_i)=(\mathcal{H}_i^{T}C_{\mb{x}_i^f,l_i}^{-1}\mathcal{H}_i)^{-1}\mathcal{H}_i^{T}C_{\mb{x}_i^f,l_i}^{-1}Y_{\mb{x}_i^{f}}(\mb{D}_M)
\end{equation} 
with $l_i$ being assumed known, $\mathcal{H}_i\in \mathcal{M}_{M,m}(\rr)$ being equal to $[H_i(\lambda_1),\cdots,H_i(\lambda_M)]^{T}$ and $C_{\mb{x}_i^f,l_i}=[C_{\mb{x}_i^f,l_i}(\lambda_s,\lambda_{s^{\prime}})]_{s,s^{\prime}}\in\mathcal{M}_{M,M}(\rr)$. The statistical uncertainty related to $\hat{\beta}_i$ is equal to
\begin{equation}
\label{beta_uncertainty}
\text{Cov}(\hat{\beta}_i(l_i))=(\mathcal{H}_i^{T}C_{\mb{x}_i^f,l_i}^{-1}\mathcal{H}_i)^{-1}
\end{equation}
If the mean is constant, i.e. $m_{\mb{x}_i^{f}}(\lambda)=\beta_i\in\rr$ for any $\lambda\in\mathcal{L}$, then Equation (\ref{beta_estimate}) and (\ref{beta_uncertainty}) become respectively
\begin{equation}
\label{beta_estimate_constant}
\hat{\beta}_i(l_i)=(\mathbf{1}^{T}_{\rr^{M}}C_{\mb{x}_i^f,l_i}^{-1}\mathbf{1}_{\rr^{M}})^{-1}\mathbf{1}^{T}_{\rr^{M}}C_{\mb{x}_i^f,l_i}^{-1}Y_{\mb{x}_i^{f}}(\mb{D}_M)
\end{equation}
and 
\begin{equation}
\label{beta_uncertainty_constant}
\text{Cov}(\hat{\beta}_i(l_i))=(\mathbf{1}^{T}_{\rr^{M}}C_{\mb{x}_i^f,l_i}^{-1}\mathbf{1}_{\rr^{M}})^{-1}
\end{equation}
Let us point out that the computation of the predictive variance (\ref{predVar}) is now modified due to the extra uncertainty coming from Equation (\ref{beta_uncertainty}). Equation (\ref{predVar}) has to be replaced by
\begin{multline}
\label{predVar_estimatedbeta}
V^{M}_{\mb{x}_i^f,\hat{\beta}_i(l_i),\sigma_i^2}(\lambda,\lambda^{\prime})=\sigma_i^{2}\Big[C_{\mb{x}_i^f,l_i}(\lambda,\lambda^{\prime})-C_{\mb{x}_i^f,l_i}\big(\lambda,\mb{D}_M\big)^{T}C_{\mb{x}_i^f,l_i}(\mb{D}_M)^{-1}C_{\mb{x}_i^f,l_i}\big(\lambda^{\prime},\mb{D}_M\big)+\\
\big(H_i(\lambda)-C_{\mb{x}_i^f,l_i}\big(\lambda,\mb{D}_M\big)^{T}C_{\mb{x}_i^f,l_i}(\mb{D}_M)^{-1}\mathcal{H}_i\big)
\big(\mathcal{H}_i^{T}C_{\mb{x}_i^f,l_i}^{-1}\mathcal{H}_i\big)^{-1}\\
\big(H_i(\lambda^{\prime})-C_{\mb{x}_i^f,l_i}\big(\lambda^{\prime},\mb{D}_M\big)^{T}C_{\mb{x}_i^f,l_i}(\mb{D}_M)^{-1}\mathcal{H}_i\big)
\Big].
\end{multline}
The proof of this equation can be found in Chapter $4$ of \citet{San03}. When the mean is constant, i.e. $H_i(\lambda)=1$ in Equation (\ref{predVar_estimatedbeta}), then
\begin{multline}
V^{M}_{\mb{x}_i^f,\hat{\beta}_i(l_i),\sigma_i^2}(\lambda,\lambda^{\prime})=\sigma_i^{2}\Big[C_{\mb{x}_i^f,l_i}(\lambda,\lambda^{\prime})-C_{\mb{x}_i^f,l_i}\big(\lambda,\mb{D}_M\big)^{T}C_{\mb{x}_i^f,l_i}(\mb{D}_M)^{-1}C_{\mb{x}_i^f,l_i}\big(\lambda^{\prime},\mb{D}_M\big)+\\
\big(1-C_{\mb{x}_i^f,l_i}\big(\lambda,\mb{D}_M\big)^{T}C_{\mb{x}_i^f,l_i}(\mb{D}_M)^{-1}\mathcal{H}_i\big)
\big(\mathcal{H}_i^{T}C_{\mb{x}_i^f,l_i}^{-1}\mathcal{H}_i\big)^{-1}\\
\big(1-C_{\mb{x}_i^f,l_i}\big(\lambda^{\prime},\mb{D}_M\big)^{T}C_{\mb{x}_i^f,l_i}(\mb{D}_M)^{-1}\mathcal{H}_i\big)
\Big].
\end{multline}
Then, the MLE of $\sigma_i^2$ knowing $l_i$ is also given explicitly by 
\begin{equation}
\hat{\sigma}_i^2(l_i)=\frac{1}{M}\Big(Y_{\mb{x}_i^{f}}(\mb{D}_M)-\mathcal{H}_i\hat{\beta}_i\Big)^{T}C_{\mb{x}_i^f,l_i}(\mb{D}_M)^{-1}\Big(Y_{\mb{x}_i^{f}}(\mb{D}_M)-\mathcal{H}_i\hat{\beta}_i\Big).
\end{equation}
By putting $\hat{\beta}_i(l_i)$ and $\hat{\sigma}_i^2(l_i)$ into the likelihood expression, then the MLE estimator of $l_i$ is obtained by solving the following minimization problem$:$
\begin{equation}
\hat{l}_i=\argmin{l_i}{\Big[\log{(\hat{\sigma}_i^2(l_i))}}+\frac{1}{M}\log{|C_{\mb{x}_i^f,l_i}(\mb{D}_M)|\Big]}.
\end{equation}

\begin{Rem}
In a recent work dealing with GP-based inverse problems \citep{Fu15}, only one GP emulator is constructed across $\mb{x}^{f}$ and $\lambda$ together. As the dimension of $\mb{x}^{f}$ increases ($\geq 10$), the accuracy of such an emulator could drop if the number of learning simulations is not large enough. However, the dimension of $\mb{x}^f$ in the COSI tests is not the problem because this vector is formed by the injection flowrate, the injection temperature, the pressure and the level of water in the cold leg. The pressure varies between $2.5$ and $70.6$ bars, including $25$ tests launched at $70.6$ bars, $10$ tests at $50$ bars, $6$ tests at $3.7$ bars, etc. Such patterns are observed for the level of water as well. The design of experiments across $(\mb{x}^{f},\lambda)\in\rr^{6}$ would therefore include a few well-covered subspaces and other empty ones. This is why we still opted for one GP emulator per each site $\mb{x}_i^f\in\mb{X}^{f}$.
\end{Rem}

\vspace{3pt}\hrule\vspace{6pt}
\subsection{The Metropolis-Hastings (MH) algorithm}

\label{MH}

\medskip
In the non-linear case, the blocked Gibbs sampler includes an inner MH step at every iteration $1\leq k\leq N_{mcmc}$. The MH algorithm relies on a proposal distribution which has to comply with some properties so that the Markov chain converges \citep{Chib95}. As the density (\ref{cond_latent}) is written as the product of the $n$ marginal densities $\pi(\alpha_i|m,\sigma^2,z_i^{f},\mb{x}_i^f)$, we just need to focus on how the MH algorithm can generate samples from $\alpha_i|m,\sigma^2,z_i^{f},\mb{x}_i^f$.

\begin{enumerate}
\item \textbf{Initialization}$:$  Start with $\alpha_i(0)$;
\item \textbf{Loop}$:$ Choose a number $N_{mh}$ of iterations, then for $1\leq s\leq N_{mh}$ generate$:$ 

\smallskip
\begin{itemize}
\item $\alpha_i(new)\thicksim \pi_{prop}\big(.|m(s),\sigma^{2}(s),\alpha_i(s)\big)$ with $\pi_{prop}(.|.)$ being a proposal density chosen by the user.
\item $\alpha_i(s+1)=\alpha_i(new)$ with probability $p_{mh,i}$ equal to
\begin{equation}
\label{probaMH}
\min{\left(1\,,\frac{\pi\big(\alpha_i(new)|m(s),\sigma^2(s),z_i^{f},\mb{x}_i^f\big)\pi_{prop}\big(\alpha_i(s)|m(s),\sigma^{2}(s),\alpha_i(new)\big)}{\pi\big(\alpha_i(s)|m(s),\sigma^2(s),z_i^{f},\mb{x}_i^f\big)\pi_{prop}\big(\alpha_i(new)|m(s),\sigma^{2}(s),\alpha_i(s)\big)}\right)}
\end{equation}
and $\alpha_i(s+1)=\alpha_i(s)$ with probability $1-p_{mh,i}$.
\end{itemize}
\end{enumerate}

\begin{thm}
\label{mhtheo}
By using a proposal density independent from the current state $\alpha_i(s)$, that is,
\begin{equation}
\label{proposal_ind}
\pi_{prop}\big(\alpha_i(new)|m(s),\sigma^{2}(s),\alpha_i(s)\big)=\pi_{prop}\big(\alpha_i(new)|m(s),\sigma^{2}(s)\big)
\end{equation} 
then the probability (\ref{probaMH}) can be simplified to
\begin{equation}
\label{probaMH_simplified}
p_{mh,i}=\min{\left(1\,,\frac{\pi\big(z_i^{f}|m(s),\sigma^2(s),\alpha_i(new),\mb{x}_i^f\big)}{\pi\big(z_i^{f}|m(s),\sigma^2(s),\alpha_i(s),\mb{x}_i^f\big)}\right)}
\end{equation}
where
\begin{equation}
\pi\big(z_i^{f}|m(s),\sigma^2(s),\alpha_i,\mb{x}_i^f\big)=\frac{1}{\sqrt{2\pi}\sigma_{\epsilon_i}}\exp{\Big\{-\frac{1}{2}\frac{\big(z_i^f-Y_{\exp{\alpha_i}}(\mb{x}_i^f)\big)^{2}}{\sigma^2_{\epsilon_i}}\Big\}}.
\end{equation}
\end{thm}

\begin{pf}
If we use the Bayes formula, the ratio
\begin{equation}
\frac{\pi\big(\alpha_i(new)|m(s),\sigma^2(s),z_i^{f},\mb{x}_i^f\big)\times\pi_{prop}\big(\alpha_i(s)|m(s),\sigma^{2}(s),\alpha_i(new)\big)}{\pi\big(\alpha_i(s)|m(s),\sigma^2(s),z_i^{f},\mb{x}_i^f\big)\times \pi_{prop}\big(\alpha_i(new)|m(s),\sigma^{2}(s),\alpha_i(s)\big)}
\end{equation}
is equal to
\begin{multline}
\label{eq2_appendix}
\frac{\pi\big(z_i^{f}|m(s),\sigma^2(s),\alpha_i(new),\mb{x}_i^f\big)\times\pi(\alpha_i(new)|m(s),\sigma^2(s))}{\pi\big(z_i^{f}|m(s),\sigma^2(s),\alpha_i(s),\mb{x}_i^f\big)\times\pi(\alpha_i(s)|m(s),\sigma^2(s))}
\times \\
\frac{\pi_{prop}\big(\alpha_i(s)|m(s),\sigma^{2}(s),\alpha_i(new)\big)}{\pi_{prop}\big(\alpha_i(new)|m(s),\sigma^{2}(s),\alpha_i(s)\big)}.
\end{multline}
Finally, as $\pi_{prop}(.|.)$ is chosen independent of $\alpha_i(.)$, then Equation (\ref{eq2_appendix}) comes down to
\begin{equation}
\frac{\pi\big(z_i^{f}|m(s),\sigma^2(s),\alpha_i(new),\mb{x}_i^f\big)}{\pi\big(z_i^{f}|m(s),\sigma^2(s),\alpha_i(s),\mb{x}_i^f\big)}.
\end{equation}
\end{pf}
By default, the proposal density (\ref{proposal_ind}) can be chosen as a Gaussian density with mean $m(s)$ and variance $\sigma^2(s)$. The variance matrix should be tuned to improve mixing efficiency of the Markov chain so that the acceptance rate is near $0.45$ when $\alpha_i(.)\in\rr$ and tends to $0.25$ as the dimension of the problem increases \citep{Roberts97}. However, as the MH algorithm is embedded within a Gibbs sampler, the convergence of (\ref{cond_latent}) itself is not of primary importance. In Section \ref{cositests}, we have chosen $N_{mh}=10$ while \citet{Muller91} even proposes $N_{mh}=1$. The interest of such a simplification is stressed in \citet{Robert98} where the authors mention that 
\begin{enumerate}
\item the stationary distribution of the Markov chain remains the same$:$ $m,\sigma^2,\mb{A}|\mb{z}^{f},\mb{X}^{f}$;
\item a more exhaustive sampling of the full conditional distribution does not necessarily lead to better estimates.
\end{enumerate}

\vspace{3pt}\hrule\vspace{6pt}
\subsection{Proof of Theorem \ref{GPmodular_fcd}}

\label{modularAppendix}

\begin{pf}
Let
\begin{equation}
Y(\mb{D}_M)=\cup_{i=1}^{n} Y_{\mb{x}_i^f}(\mb{D}_M)
\end{equation}
be the union of the learning simulations $Y_{\mb{x}_i^f}(\mb{D}_M)$ for fitting the $i$-th GP emulator and let 
\begin{equation}
\theta=\cup_{i=1}^{n}\{\theta_i \}
\end{equation}
be the set of all hyperparameters $\theta_i=(\beta_i,\sigma_i^2,l_i)$ related to the $i$-th GP emulator constructed at $\mb{x}_i^f$ ($1\leq i\leq n$).
If a full Bayes approach is applied, then $\theta$ is estimated jointly with $(m,\sigma^2)$ and the posterior distribution of $(m,\sigma^2,\theta)$ is not only conditional on $\mb{z}^f$ but also on $Y(\mb{D}_M)$. Hence,
\begin{equation}
\label{joint_post_gp_cond}
\pi(m,\sigma^{2},\theta,\mb{A}|\mb{z}^{f},\mb{X}^{f},Y(\mb{D}_M))\propto \mathcal{L}(\mb{z}^{f},Y(\mb{D}_M)|m,\sigma^{2},\theta,\mb{A},\mb{X}^{f})\pi(\mb{A}|m,\sigma^{2},\theta)\pi(m,\sigma^{2},\theta).
\end{equation}
where 
\begin{equation}
\mathcal{L}(\mb{z}^{f},Y(\mb{D}_M)|m,\sigma^{2},\theta,\mb{A},\mb{X}^{f})=\prod_{i=1}^{n}\mathcal{L}_i(z_i^{f},Y_{\mb{x}_i^f}(\mb{D}_M)|m,\sigma^{2},\theta_i,\alpha_i,\mb{x}_i^f).
\end{equation}
Every one dimensional likelihood $\mathcal{L}_i(z_i^{f},Y_{\mb{x}_i^f}(\mb{D}_M)|m,\sigma^{2},\theta_i,\alpha_i)$ is a Gaussian density with a mean equal to
\begin{equation}
(m_{\mb{x}_i^f,\beta_i}(\exp{\alpha_i}), m_{\mb{x}_i^f,\beta_i}(\mb{D}_M))^{T}\in\rr^{M+1}
\end{equation}
and a covariance matrix equal to
\begin{equation}
\sigma_i^2
\begin{pmatrix}
\vspace{0.2cm}
C_{\mb{x}_i^f,l_i}(\exp{\alpha_i},\exp{\alpha_i})+\frac{\sigma_{\epsilon_i}^2}{\sigma_i^{2}} & C_{\mb{x}_i^f,l_i}\big(\exp{\alpha_i},\mb{D}_M\big)\\
C_{\mb{x}_i^f,l_i}\big(\exp{\alpha_i},\mb{D}_M\big)^T & C_{\mb{x}_i^f,l_i}(\mb{D}_M).
\end{pmatrix}\, \in \mathcal{M}_{M+1,M+1}(\rr)
\end{equation}
Such a full-Bayes approach requires specifying a prior distribution $\pi(\theta)$, which can be complicated for the correlation lengths \citep{Paulo05}. In addition, we can see that the physical experiments $\mb{z}^{f}$ impact the estimation of $\theta$ whereas we would like that only the learning simulations $Y_{\mb{x}_i^f}(\mb{D}_M)$ are used for it. Both remarks conducted us to turn to another approach. 

\medskip
Indeed, Equation (\ref{cond_latent_pg}) is derived from applying \textit{a modular approach} \citep{Liu09} in the same spirit than some papers dealing with GP-based inverse UQ \citep{Koh2001,Bayarri2007,Fu15}. 
Such a technique neglects the impact of the physical experiments $\mb{z}^{f}$ on $\theta_i$ by estimating a point estimate $\hat{\theta}_i$ with respect to $Y_{\mb{x}_i^f}(\mb{D}_M)$ only (see \ref{GPappendix}), then putting it into the conditionnal distribution $\mathcal{L}^{Cond}$ of $\mb{z}^f$ conditional on $m$, $\sigma^2$, $Y(\mb{D}_M)$ and $\mb{A}$$:$
\begin{equation}
\mathcal{L}^{Cond}(\mb{z}^{f}|Y(\mb{D}_M),\mb{A},\mb{X}^{f},m,\sigma^2,\hat{\theta})=\prod_{i=1}^{n} \mathcal{L}^{Cond}_i(z_i^f|Y_{\mb{x}_i^f}(\mb{D}_M),\alpha_i,m,\sigma^2,\hat{\theta}_i)
\end{equation}
where each $\mathcal{L}^{Cond}_i(.|.)$ is the density of a Gaussian distribution with mean equal to
\begin{equation}
\mu^{M}_{\mb{x}_i^f,\hat{\theta}_i}(\exp{\alpha_i})
\end{equation}
and variance equal to
\begin{equation}
V^{M}_{\mb{x}_i^f,\hat{\theta}_i}(\exp{\alpha_i},\exp{\alpha_i})+\sigma_{\epsilon_i}^2
\end{equation}
with $\mu^{M}_{\mb{x}_i^f,\hat{\theta}_i}(.)$ and $V^{M}_{\mb{x}_i^f,\hat{\theta}_i}(.)$ being respectively the predictive mean and variance of the $i$-th GP emulator defined at $\hat{\theta}_i$. Finally, Equation (\ref{cond_latent_pg}) is obtained as the conditional posterior distribution of $\mb{A}$ by replacing $\mathcal{L}(\mb{z}^{f},Y(\mb{D}_M)|\mb{A},\mb{X}^{f},m,\sigma^2,\theta)$ with $\mathcal{L}^{Cond}(\mb{z}^{f}|Y(\mb{D}_M),\mb{A},\mb{X}^{f},m,\sigma^2,\hat{\theta})$ in Equation (\ref{joint_post_gp_cond}).


\end{pf}

\subsection{Cost of the GP-based Gibbs sampler in terms of CPU time}

\label{GP_computation_cost}

\medskip
The main expense in computing both the predictive mean and variance given by Equations (\ref{predMean}) and (\ref{predVar_estimatedbeta}) is to compute the inverse of the correlation matrix at $\mb{D}_M$, which requires $\mathcal{O}(M^{3})$ operations. The modular approach actually mitigates it because the inversion is needed only once before the Gibbs sampler is started up. Then, at each iteration of the sampler, the calculation of the predictive mean and variance costs $\mathcal{O}(M)$ and $\mathcal{O}(M^{2})$ respectively. The overall CPU time needed by the GP-based sampler thus increases with $M$. The larger $M$, the more accurate the GP approximations, the closer the resulting GP-based posterior distribution is expected to the exact posterior distribution (\ref{post_expression_nonlinear}). A trade-off between the size of $\mb{D}_M$ and the CPU time of the Gibbs sampler is then to be found. Methods such as local GP  approximations \citep{Gram15} or nested GP \citep{Rulliere17} could help to speed up the sampler for large $\mb{D}_M$ ($M$ equal to several thousand).

\section{Convergence diagnostics for MCMC samplers}

\label{MCMCconv}

The convergence of the Gibbs samplers implemented in the paper was checked in three ways as advised by \citet{Robert98}$:$
\begin{enumerate}
\item \underline{convergence of the Markov chain to the stationary distribution} to check whether the distribution of the chain remains unchanged as iterations $k$ increase.
\item \underline{convergence of averages}, to assess the precision of posterior estimates, for example:
\begin{equation}
\label{MCMC_mean_estimate}
\hat{b}=\frac{1}{N_{mcmc}}\sum_{k=1}^{N_{mcmc}} b(k)
\end{equation}
Unlike regular Monte Carlo (MC) estimators based on i.i.d. samples, the $\{b(k)\}_k$ are no longer independent. Such MCMC estimators still converge to the posterior mean of $b$, but they are less precise than MC ones. The convergence of averages deals with choosing $N_{mcmc}$ so that point estimates such as (\ref{MCMC_mean_estimate}) achieve a prescribed level of precision.    
\item \underline{convergence to i.i.d. sampling}, where we seek to pick up MCMC samples that are as independent as possible. 
\end{enumerate}
A stationary test can be implemented for the first objective. For the second objective, the so-called \textit{effective sample size} (e.s.s.) is a useful indicator providing the number $N_{mc}<N_{mcmc}$ of hypothetical i.i.d. samples to achieve the same accuracy than (\ref{MCMC_mean_estimate}). The e.s.s. is calculated from autocorrelation between posterior samples. The smaller the autocorrelation, the larger the e.s.s.. 

The third goal is called sub-sampling which is to pick up quasi-independent samples from the chain.

\medskip
The CODA library of the R statistical software implements various indexes to monitore the convergence of MCMC sampling \citep{Coda}. For a comprehensive comparison of convergence diagnostics, see for instance \citet{Cowles96}.

\paragraph{Geweke test} The principle of this test is to compare the first $10\%$ samples of the chain with the last $50\%$ samples in terms of mean estimate. If the Z-test for mean comparison is passed, then it is likely that the chain reaches its stationary distribution somewhere in the first $10\%$. The remaining $90\%$ samples are then judged to be generated from the posterior distribution.

\paragraph{The Gelman and Rubin's statistic}
We have computed this statistic in Section \ref{nonlin_cosi} to diagnose any potential issue of non convergence to the posterior distribution \citep{GelmanRubin92}. It computes the ratio of the between-chains variance to the within-chain variance for each parameter $m_1$, $m_2$, $\sigma_1^2$, $\sigma_2^2$. A rule of thumb to declare convergence is that the ratio is below $1.10$, but more generally the closest the ratio to $1$, the most likely the convergence. In Section \ref{nonlin_cosi}, the ratio dropped below $1.05$ as iterations increase, except in Setting 1 where the ratio computed for $\sigma_2^2$ rose to $1.09$ after $10^{6}$ iterations, then decreased again to $1$ beyond $2\times 10^{6}$ iterations. This should be due to poor mixing of samples.

\section{Difference between CIRCE and calibration of parameters}

\label{Calib_vs_CIRCE}

\medskip
The CIRCE method comes down to conducting probabilistic inversion. Based on a probability distribution of the differences between the simulations and the experimental QoIs, CIRCE aims to get back to the probability distribution of uncertain inputs, such as multiplicative factors in the paper. Another example of probabilistic inversion can for example be found in \citet{Fu15} where the probability distribution of Strickler coefficients is estimated in a flood risk problem. Although probabilistic inversion and calibration are both part of inverse UQ methods, the former aims to recover a probability distribution whereas the latter infers best-fitting parameters. If calibration were performed instead of CIRCE, Equation (\ref{CIRCEmodel_short}) would be replaced by
\begin{equation}
z_i^{f}=\, Y_{\lambda}(\mb{x}_i^{f})+\epsilon_i
\end{equation}
in which the realization $\lambda$ of the multiplicative factor(s) would be the same for every $\mb{x}_i^f\in\mb{X}^{f}$, as done in \citet{Wu18b}. 



\end{document}